\documentclass [12pt, letterpaper]{article}
\pdfoutput=1
\usepackage{fullpage}
\usepackage{amsmath}
\usepackage{amssymb}
\usepackage{graphicx}
\usepackage{mathrsfs}
\usepackage{wrapfig}
\usepackage{boxedminipage}
\usepackage{epsfig}
\usepackage{setspace}
\usepackage{subfigure}
\usepackage{float}
\newcommand{\reef}[1]{(\ref{#1})}
\DeclareSymbolFont{AMSb}{U}{msb}{m}{n}
\DeclareMathSymbol{\IN}{\mathbin}{AMSb}{"4E}
\DeclareMathSymbol{\IZ}{\mathbin}{AMSb}{"5A}
\DeclareMathSymbol{\IR}{\mathbin}{AMSb}{"52}
\DeclareMathSymbol{\Q}{\mathbin}{AMSb}{"51}
\DeclareMathSymbol{\II}{\mathbin}{AMSb}{"49}
\DeclareMathSymbol{\IC}{\mathbin}{AMSb}{"43}
\DeclareMathSymbol{\IP}{\mathbin}{AMSb}{"50}
\DeclareMathSymbol{\IH}{\mathbin}{AMSb}{"48}
\DeclareMathSymbol\IA{\mathalpha}{AMSb}{"41}
\DeclareMathSymbol\IS{\mathalpha}{AMSb}{"53}

\def\Q{{\cal Q}}

\begin{document}

\begin{flushright}
\phantom{{\tt arXiv:0808.????}}
\end{flushright}

\bigskip
\bigskip
\bigskip

\begin{center} 

{\Large \bf String Theory and   Water Waves}  

\end{center}

\bigskip \bigskip \bigskip \bigskip

\centerline{\bf Ramakrishnan Iyer$^\natural$, Clifford V. Johnson$^\flat$, Jeffrey S. Pennington$^\sharp$}

\bigskip
\bigskip
\bigskip

  \centerline{\it $^{\natural,\flat}$Department of Physics and Astronomy }
\centerline{\it University of
Southern California}
\centerline{\it Los Angeles, CA 90089-0484, U.S.A.}

\bigskip

  \centerline{\it $^\sharp$SLAC National Accelerator Laboratory}
\centerline{\it Stanford University}
\centerline{\it Stanford, CA 94309, U.S.A.}

\bigskip

\centerline{\small \tt $^\natural$ramaiyer, $^\flat$johnson1,  [at] usc.edu; $^\sharp$jpennin [at] stanford.edu}

\bigskip
\bigskip


\begin{abstract} 
\noindent 
We uncover a remarkable role that an infinite hierarchy of non--linear
differential equations plays in organizing and connecting certain
${\hat c}<1$ string theories non--perturbatively. We are able to embed
the type 0A and 0B $(A,A)$ minimal string theories into this single
framework. The string theories arise as special limits of a rich
system of equations underpinned by an integrable system known as the
dispersive water wave hierarchy. We observe that there are several
other string--like limits of the system, and conjecture that some of
them are type~IIA and~IIB $(A,D)$ minimal string backgrounds. We
explain how these and several string--like special points arise and
are connected.  In some cases, the framework endows the theories with
a non--perturbative definition for the first time. Notably, we
discover that the Painlev\'e IV equation plays a key role in
organizing the string theory physics, joining its siblings, Painlev\'e
I and II, whose roles have previously been identified in this minimal
string context.
\end{abstract}
\newpage \baselineskip=18pt \setcounter{footnote}{0}

\section{Introduction}
\label{sec:introduction}

While string theory has had remarkable successes over the last several
years, accelerated by the revolutions in understanding its
non--perturbative properties, it is still very much the case that we
do not yet know what the theory is. We cannot state unambiguously what
the basic degrees of freedom are (it is highly context dependent in a
way that depends upon the dynamics themselves), and even the
backgrounds in which the theory propagates are themselves open to
interpretation. For example, in some descriptions and situations, the
theory contains gravity, and in others, it does not. From some
perspectives there are open strings present, and from others, only
closed. Ironically, several of these frustrating (from the point of
view of finding simple definitions) features are also among the
theory's most powerful positive traits, allow an ever--widening range
of applications of the theory to diverse problems, often of a strongly
coupled nature.

While applications continue, it is still important to try to get to
grips with what the theory is. At the very least, this is important
from a pragmatic standpoint, since perhaps a useful definition or
characterization of string theory might be put to use as, for example,
a diagnostic device in identifying when a physical problem may have
some aspect of it that is amenable to solution by string theory
methods.  More generally, if string theory ultimately plays some
fundamental role in the understanding of physics beyond the standard
model, and/or in cosmology and other origins questions about the
universe at large, a more profound understanding of the nature, power,
and scope of the theory would seem to be highly desirable.

At best, to date, as a result of various dualities, we know that it is
probably part of some larger physical framework which itself is only
string theoretic in various corners of its parameter space. This
physical setting, called M--theory, remains profoundly mysterious well
over a decade after the first clear glimpses of
it\cite{Hull:1994ys,Townsend:1995kk,Witten:1995ex}.
 
Historically, problems pertaining to such essential matters of
understanding in physics are greatly illuminated by having a rich set
of examples that are simple, but yet complex enough to contain all the
important phenomena in question. For the problems outlined above, it
would be rather excellent to have the simplest possible string
theories that still contain some of the marvellous non-perturbative
physics we know and love, and  be able to follow them as they
connect to each other in ways that are entirely invisible in
perturbation theory.  Further icing on the cake would be to have  the
physics  all captured in terms of relatively familiar structures
for which there is an existing technology for its study.

This is the subject of this paper (and a follow--up to appear
later\cite{companion}), at least in part. The simplest known strings
with tractable non--perturbative physics that contain a rich set of
phenomena (such as holography and open--closed dualities) are the
minimal strings\cite{Gross:1989vs,Brezin:1990rb,Douglas:1989ve}, and
in particular (where non--perturbative physics is concerned) the
type~0 strings (formulated in
refs.\cite{Morris:1990bw,Dalley:1991qg,Dalley:1991vr,Johnson:1992pu,Dalley:1992br}
and refs.\cite{Crnkovic:1990ms,Hollowood:1991xq}, and recognized as
type~0 strings in ref.\cite{Klebanov:2003wg}). The non--perturbative
formulation of the type 0A and type 0B strings can be done rather
beautifully in terms of certain integrable systems, as we will review
later: Type 0A has the Korteweg--de Vries system while type 0B has
Zakharov--Shabat. The non--perturbative physics of each is formulated
in terms of associated non--linear ordinary differential equations
often called ``string equations''. While much of the language of the
two is similar, these are very different systems, and except for
various accidental (from the perspective of those separate
formulations) perturbative coincidences (and a non--trivial
non--perturbative equivalence for one model --- see later), the
physics of each are quite separate indeed.

This paper builds on all of these results, taking them much
further. We have found that there is a larger framework into which the
type 0A and type 0B string theories can be naturally embedded and
within which they are connected as parts of a larger theory. We found
further that the two string theories are merely special points in a
much larger tapestry of possibilities. When perturbation theory is
examined, other special points suggest themselves, and they turn out
to be just as ``stringy'' as the original type~0 theories, deserving
to be thought of as string theories as well. We begin the program of
trying to identify some of these theories, with some success. We also
find that the larger framework provides natural definitions of regimes
of the type~0 string theories that are hard to define using
perturbation theory, and we will report more fully on
non--perturbative aspects in a follow--up paper\cite{companion}.

In this sense, we have a precise analogue of M--theory. We have a
larger physical framework that is not itself a string theory, but that
can be readily specialized to yield string theories as special
limits. We can move between different theories in a quite natural way,
which is nonetheless outside the framework of any of its daughter
string theories. We find this encouraging and exciting.

At the base of our infinite family of string equations, organizing
much of this remarkable structure, is a non--linear differential
equation known as Painlev\'e~IV. This well--known equation from the
classical mathematics literature\footnote{See for example the lovely
  monograph of ref.\cite{Noumi:2004} and references therein.}, part of
a celebrated family of six equations, has two arbitrary constants,
usually denoted $\alpha$ and $\beta$. (Actually, two copies of
Painlev\'e IV turn up in our story, intertwined in an interesting
way.) It turns out that the type 0A and 0B points in the tapestry of
theories occur at the vanishing of one or other of these constants for
one of the copies of Painlev\'e~IV.  The vanishing of the constants of
the second Painlev\'e IV hint at interesting new special points.

After reviewing crucial aspects of the type~0A and~0B string theories
in section~\ref{sec:type0}, we unpack the dispersive water wave
hierarchy and present the infinite family of equations we propose as
the string equations in
section~\ref{sec:dww}. Section~\ref{sec:painleve} highlights the role
of Painlev\'e~IV. In section~\ref{sec:connectAB} we show how the
structures of section~\ref{sec:type0} arise as special points in this
larger framework, while section~\ref{sec:beyond} is a detailed study
of the rich properties of the string equations and the types of
solutions available. We organize and classify a great deal of the
physics that appears, and notice in section~\ref{sec:square} that much
of the physics can be organized in terms of a square. The square is
reminiscent of the main square organizing the moduli space of ${\hat
  c}=1$ (two--dimensional) string theories, discovered in
ref.\cite{Seiberg:2005bx}, and we contemplate a possible connection,
perhaps induced by Renormalization Group
flow\cite{Gross:1990ub,Hsu:1992cm} to the ${\hat c}<1$ context of the
work in question. The possible relation between the squares helps us
make a conjecture about the nature of two new special points we find:
They might be type~IIA and~IIB minimal string theories. (Note that
these are type~II theories in the sense of the structure of the GSO
projection used to formulate them. There is no spacetime
supersymmetry\cite{Seiberg:2005bx}.) In section~\ref{sec:new} we carry
out a comparison of the new structures we found to some continuum
computations for one--loop partition functions. From this we
strengthen aspects of our type~II suggestion. We conclude in
section~\ref{sec:conclusion} with a brief summary and discussion.

\section{The $(A,A)$ Type~0 Theories: Review}
\label{sec:type0}
We will start with a brief review of the type~0 string theories
coupled to the $(2,4k)$ superconformal minimal models and show how
these theories can be elegantly described within the framework of an
integrable hierarchy of partial differential equations (PDEs)
accompanied by an hierarchy of ordinary differential equations
(ODEs). These models exhibit novel and interesting physics, all of
which and more will be seen to be embedded in the DWW system that we
will describe in the next section. This review will help establish our
notation and the framework upon which we can readily build the more
general structure.

\subsection{Type 0A Strings}
\label{sec:0A}
We begin with the following ordinary differential equation (known in
the old days as a ``string equation'')
\begin{equation}\label{streqn0A}
w\mathcal{R}^2 - \frac{1}{2}\mathcal{R}\mathcal{R}'' + \frac{1}{4} \mathcal{R}'^{2} = \nu^2 \Gamma^2 \quad .
\end{equation}
This equation (or, really, family of equations) and its properties
have been studied in several papers. It was first derived and studied
as a fully non--perturbative definition of a string theory in
refs.\cite{Dalley:1991qg,Dalley:1991vr,Johnson:1992pu,Dalley:1992br},
and evidence that it defines a type~0A string theory was presented
first in ref\cite{Klebanov:2003wg}. Further properties of the
equation, in particular concerning how branes and fluxes are encoded
by it and the underlying integrable (KdV) system, were presented in
refs.\cite{Carlisle:2005mk,Carlisle:2005wa}. Here $w(z)$ is a real
function of the real variable $z$, a prime denotes $\nu
{\partial}/{\partial z}$, and $\Gamma$ and $\nu$ are real constants.
The quantity $\mathcal R$ is defined by
\begin{equation}\label{GDpoly}
\mathcal{R} = \sum_{k = 0}^{\infty} \Big ( k + \frac{1}{2}\Big ) t_k P_k \quad ,
\end{equation}
where the $P_k[w]$ are polynomials in $w(z)$ and its $z$--derivatives,
called the {Gel'fand--Dikii} polynomials\cite{Gelfand:1975rn}. They are
related by a recursion relation (defining a recursion operator $R_2$)
\begin{equation}\label{KdVrec}
P'_{k+1} = \frac{1}{4}P'''_{k} - wP'_{k} - \frac{1}{2}w'P_{k}\equiv R_2 P'_k\ , 
\end{equation}
and fixed by the value of the constant $P_{0}$ and the requirement
that the rest vanish for vanishing $w$. Some of them are:
\begin{eqnarray}\label{0APolys}
P_{0} = \frac{1}{2}; \quad P_{1} = -\frac{1}{4} w; \quad P_{2} = \frac{1}{16}(3 w^2 - w''); \nonumber\\
P_{3} = -\frac{1}{64}(10w^3 - 10ww'' - 5(w')^2 + w'''); \cdots
\end{eqnarray}
The $k$th model is chosen by setting all the $t_j$ to zero except
$t_{0} \equiv z$ and
\begin{equation}
t_{k}=\frac{(-4)^{k+1}(k!)^2}{(2k+1)!}\ .
\end{equation}
This number is chosen so that the coefficient\footnote{This gives $w=z^{1/k} +\ldots$ as $z\rightarrow+\infty$. If we had instead chosen $t_0=-z$,
we would have chosen the coefficient of $w^k$ to be unity.} of $w^k$ in $\mathcal{R}$ is set
to $-1$.

The function $w(z)$ defines the partition function $Z = \exp (-F)$ of
the string theory $via$
\begin{equation}\label{0AFreeEnergy}
w(z) = 2 \nu^2 \frac{\partial^2 F}{\partial \mu^2}\Big{|}_{\mu = z} \quad ,
\end{equation}
where $\mu$ is the coefficient of the lowest dimension operator in the
world--sheet theory. So $w(z)$ is a two--point function of the theory.

From the point of view of the $k$th theory, all the other $t_{j}$
represent couplings of closed string operators $\mathcal{O}_j$. It is
well known\cite{Douglas, Banks} that the insertion of each operator
 is captured in terms of the integrable KdV hierarchy of flows describing how $w(z,t_j)$ evolves in $t_j$:
\begin{equation}\label{KdVflow}
\frac{\partial w}{\partial t_{j}} = P'_{j+1} =R_2  P'_j\quad .
\end{equation}

For the $k$th model, equation (\ref{streqn0A}), which has remarkable
properties\cite{Carlisle:2005mk, Carlisle:2005wa}, is known to furnish
a complete non--perturbative definition of a family of spacetime
bosonic string theories \cite{Dalley:1992br}. The models are actually
type 0A strings \cite{Klebanov:2003wg}, based upon the $(2,4k)$
superconformal minimal models coupled to super--Liouville theory. As
superconformal theories, they have central charge
\begin{equation}
\label{eq:central}
{\hat c}=1-\frac{(2k-1)^2}{k}\ .
\end{equation}
The asymptotic expansions for the first two $k$ are:

\medskip

\noindent
{$k = 1$}
\begin{eqnarray}\label{0Aexpnsk=1}
w(z) &=& z + \frac{\nu \Gamma}{z^{1/2}} - \frac{\nu^2 \Gamma^2}{2 z^2} + \frac{5}{32} \frac{\nu^3}{z^{7/2}}\Gamma\left(4\Gamma^2 + 1\right) +\cdots  \quad (z \rightarrow \infty) \\
w(z) &=& 0 + \frac{\nu^2 (4 \Gamma^2 - 1)}{4 z^2} + \frac{\nu^4}{8} \frac{(4\Gamma^2 - 1)(4 \Gamma^2 - 9)}{z^5}+ \cdots \quad (z \rightarrow -\infty) \nonumber
\end{eqnarray}

\noindent
{$k = 2$}
\begin{eqnarray}\label{0Aexpnsk=2}
w(z) &=& z^{1/2} + \frac{\nu \Gamma}{2 z^{3/4}} - \frac{1}{24}\frac{\nu^2}{z^2}\left(6 \Gamma^2 + 1\right) + \cdots \quad (z \rightarrow \infty) \\
w(z) &=& (4 \Gamma^2 - 1)\left(\frac{\nu^2}{4 z^2} + \frac{1}{32}\frac{\nu^6}{z^7}(4 \Gamma^2 - 9)(4 \Gamma^2 - 25) + \cdots\right) \quad (z \rightarrow -\infty) \nonumber
\end{eqnarray}

\begin{figure}[ht]
  \begin{center}
  \includegraphics[width=120mm]{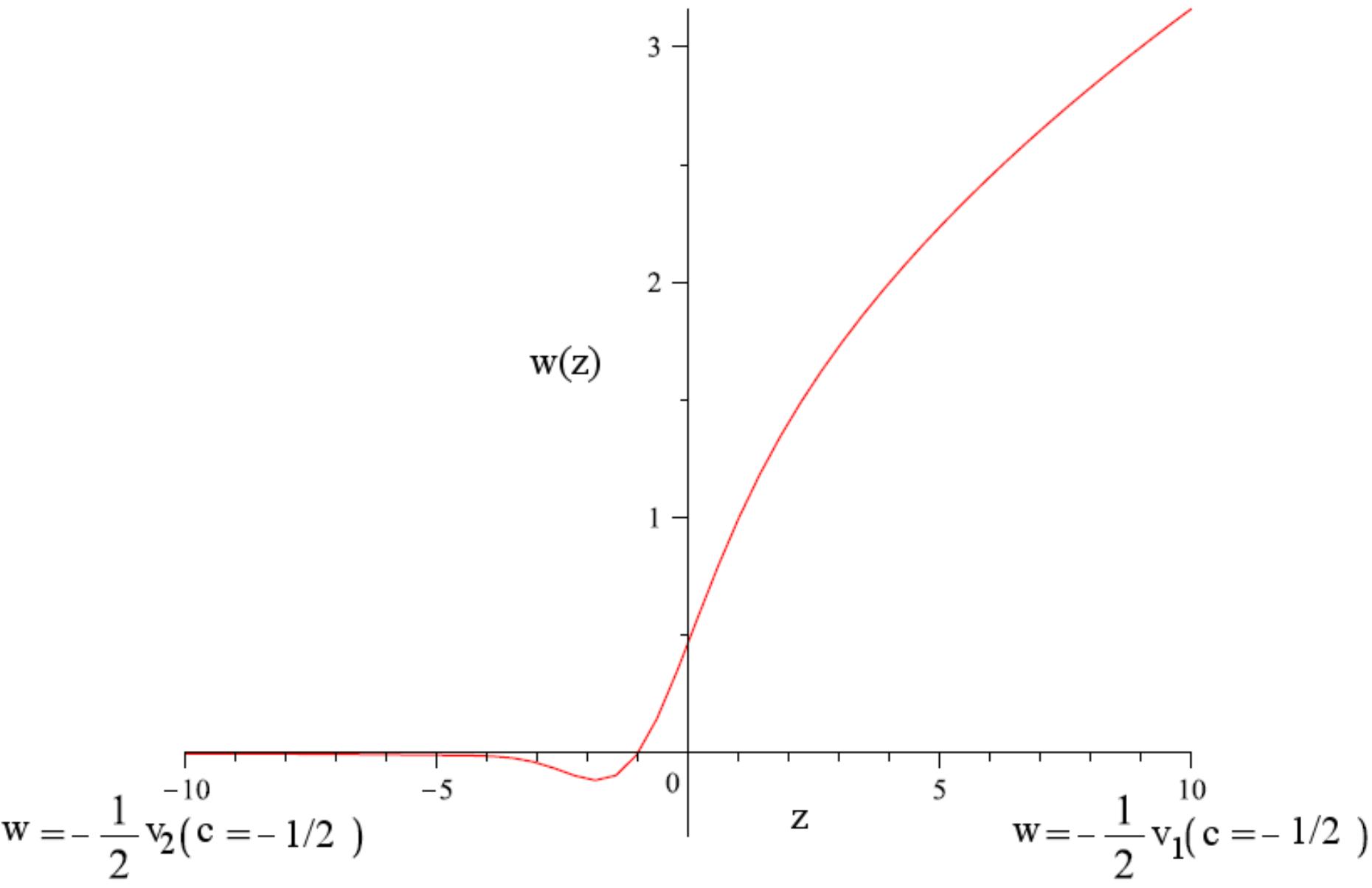}\\
  \caption{\small A plot of the $k=2$ type~0A solution showing how the
    perturbative regimes at large $|z|$ are smoothly connected.
    Section~\ref{sec:dww} discusses a function $v(x)$ ($x\propto z$),
    which has a number of different classes of behaviour distinguished
    by choice of boundary condition. The type~0A theory has class
    $v_1(z)$ in the $+z$ perturbative regime and class $v_2(z)$ in the
    $-z$ perturbative regime. Here we have set $\nu=1$ and
    $\Gamma=0$.}\label{fig:plot}
  \end{center}
\end{figure}

It should be noted that the solution for $z > 0$ can be numerically
and analytically shown to match onto the solution for $z<0$, providing
a unique\cite{Dalley:1991vr,Johnson:1992pu,Dalley:1992br}
non--perturbative completion of the theory. (See figure~\ref{fig:plot}
for an example of a solution found using numerical methods.)

As instructed in equation~\reef{0AFreeEnergy}, integrating twice the
asymptotic expansions (such as those in equations~\reef{0Aexpnsk=1}
and~\reef{0Aexpnsk=2}) furnishes the free energy $F(\mu)$, and it can
be seen to define a perturbative expansion in the dimensionless string
coupling
\begin{equation}\label{0Astrcoupling}
g_{s} = \frac{\nu}{\mu^{1 + \frac{1}{2k}}} \quad .
\end{equation}

For all models, in the $\mu \rightarrow +\infty$ regime, $\Gamma$
represents\cite{Dalley:1992br,Klebanov:2003wg} the number of
background ZZ D--branes \cite{Zamolodchikov:2001ah} in the model, with
a factor of $\Gamma$ for each boundary in the worldsheet
expansion. These are point--like branes localized at infinity in the
Liouville direction $\phi$, deep in the strong coupling region. In the
$\mu \rightarrow -\infty$ regime, $\Gamma$ represents the number of
units of RR--flux in the background, with $g_s^2 \Gamma^2$ appearing
when there is an insertion of pure RR--flux
\cite{Klebanov:2003wg}. Since there is a unique non--perturbative
solution connecting the two regimes, the string equation
(\ref{streqn0A}) supplies a non-perturbative completion of the theory
that is a very clear example of a geometric transition between these
two distinct (D-branes $vs$ RR--fluxes) spacetime descriptions of the
physics.

The function $w(z)$ is the potential in the Hamiltonian ${\cal H}\equiv
-\nu^2\partial^2_z + w(z)$ of the well--known (in the inverse
scattering literature) associated Sturm--Liouville problem connected
to the integrable KdV hierarchy. The wavefunctions of that problem
define the partition functions of
FZZT\cite{Fateev:2000ik,Teschner:2000md} D--branes stretched along the
Liouville direction $\phi$, ending at a finite $\phi_c$ set by the
eigenvalue. The zero--energy problem is
interesting\cite{Carlisle:2005mk,Carlisle:2005wa}, since there the
FZZT D--branes stretch to infinity, and the Hamiltonian's
factorization, ${\cal H}=-(\nu\partial_z\pm g(z))(\nu\partial_z\mp
g(z))$ where $w(z)=g(z)^2\pm g(z)^\prime$, is highly convenient. The
function $g(z)$ (its definition in the equation before is termed a
Miura map in the integrable literature) satisfies\cite{Dalley:1992br}
an infinite hierarchy of equations sometimes called the Painlev\'e~II
hierarchy since the equation at $k=1$ is the Painlev\'e~II
equation\footnote{Those equations were derived in a string theory
  context by studying unitary matrix
  models\cite{Periwal:1990gf,Periwal:1990qb}. Painlev\'e~II
  hierarchies have a mathematical life independent of this physical
  context, however. See {\it e.g.,} refs\cite{airault,clarkson}.}. The
asymptotic expansion of $y(z)$ generated by these equations is in
terms of worldsheets involving ZZ D--branes (or fluxes) and FZZT
D--branes. The entire problem defines a toy supersymmetric quantum
mechanics problem within which the celebrated B\"acklund
transformations of the KdV system can be made manifest. In this
language the ZZ D--branes are identified with the number of threshold
bound states (formally, zero--velocity solitons) of the system, and
the B\"acklund transformations change their number by an
integer. These more recently established
features\cite{Carlisle:2005mk,Carlisle:2005wa}, together with the
earlier identification\cite{Douglas, Banks} of the role of the KdV
flows in organizing the close string operators, show how the
integrable model and inverse--scattering technology of the
mathematical physics literature comes to life in organizing the open
and closed string content of minimal  string theory.

\subsection{Type 0B Strings}
\label{sec:0B}
Type 0B string theory coupled to the $(2, 4k)$ superconformal minimal
models \cite{Klebanov:2003wg} is described succinctly by the following
{string} equations\cite{Crnkovic:1990ms,Hollowood:1991xq}:
\begin{eqnarray}\label{streqn0B}
\sum_{l = 0}^{\infty} t_{l}(l + 1)R_{l} = 0 \ ,\qquad
\sum_{l = 0}^{\infty}t_{l}(l + 1)H_{l} + \nu q = 0 \ ,
\end{eqnarray}
where the $R_l$ and $H_l$ are polynomials of functions $r(x)$ and
$\omega (x)$ (and their derivatives), and~$\nu$ and $q$ are real
constants. 

The differential polynomials satisfy the following recursion relations
\begin{eqnarray}\label{ZSrec}
R_{l + 1} = \omega R_{l} -\left(\frac{H_l^\prime}{r}\right)^\prime+ rH_{l}\ ,\qquad
H_{l + 1}' = \omega H_{l}' - r R_{l}' \ ,
\end{eqnarray}
where a prime denotes $\nu {\partial}/{\partial x}$. Some of them are:
\begin{eqnarray}\label{ZSpoly}
H_{-1} &=& 1, \quad R_{-1} = 0;\nonumber\\
H_{0} &=& 0, \quad R_{0} = r;\nonumber\\
H_{1} &=& -\frac{r^2}{2}, \quad  R_{1} = \omega r;\nonumber\\
H_{2} &=& -r^2\omega, \quad R_{2} = -\frac{r^3}{2} + r\omega^2 + r'' ; \\
H_{3} &=& \frac{3}{8}r^4 -\frac{3}{2}r^2\omega^2 + \frac{1}{2}r'^2 - rr''  \quad ,\nonumber\\
R_{3} &=&-\frac{3}{2}r^3\omega + r\omega^3 + 3r'\omega' + 3\omega r'' + r\omega''  \quad .\nonumber
\end{eqnarray}
The function ${\widetilde w}(x) = {r^2}/{4}$ defines the partition function of the
theory $via$
\begin{equation}\label{0BFreeEnergy}
 {\widetilde w}(x) = \frac{r^2}{4} =  \nu^2 \frac{d^2F}{dx^2}  \quad .
\end{equation}
The $n$th model is chosen by setting all $t_l$ to zero except $t_0
\sim x$ and $t_n$, analogous to what was done in the previous section
concerning the 0A case. Note that these models have an interpretation
as type~0B strings coupled to the $(2,2n)$ superconformal minimal
models only for even\footnote{For odd $n$, lack of modular invariance
  of the partition function rules out the interpretation as type~0B
  strings coupled to superconformal matter \cite{Klebanov:2003wg}.}
$n$. Writing $n = 2k$, we again have a set of models connected to
the $(2,4k)$ superconformal minimal models, this time type~0B.

As in the 0A case, from the point of view of the $k$th theory, all the
other $t_j$ represent coupling to closed string operators
$\mathcal{O}_j$. Again the insertion of each operator can be expressed
in terms of the Zakharov--Shabat\cite{Zakharov:1979zz} hierarchy of flows, the underlying integrable system in this case:
\begin{eqnarray}\label{0Bflow}
\frac{\partial \beta}{\partial t_k} =  R_{k+1}\ ,\qquad
\frac{\partial r}{\partial t_k} =  -\frac{H^{'}_{k+1}}{r} \ ,
\end{eqnarray}
where $\beta^{'} \equiv \omega$.

The asymptotic expansions of the string equations (\ref{streqn0B}) for
the first  even $n=2k$ are:

\noindent
$n = 2 \,\,\, (k=1)$
\begin{eqnarray}\label{0Bexpnsm=2}
{\widetilde w}(x) &=& \frac{x}{4} + \left(q^2 - \frac{1}{4}\right)\left[\frac{\nu^2}{2x^2} + \left(q^2 - \frac{9}{4}\right)\left(\frac{-2\nu^4}{x^5} + \cdots\right)\right]\ , \quad (x \rightarrow \infty) \nonumber\\
{\widetilde w}(x) &=&  \frac{\nu q\sqrt{2}}{4|x|^{1/2}} - \frac{\nu^2 q^2}{4 |x|^2} + \frac{\nu^3}{|x|^{7/2}}\frac{5\sqrt{2}}{64} q\left(1 + 4q^2\right)+\cdots  \quad (x \rightarrow -\infty)
\end{eqnarray}

\noindent
$n = 4 \,\,\, (k=2)$
\begin{eqnarray}\label{0Bexpnsm=4}
{\widetilde w}(x) &=& \frac{\sqrt{x}}{4} + \frac{\nu^2}{144 x^2} \left(64q^2 - 15\right) + \cdots ; \quad (x \rightarrow \infty) \\
{\widetilde w}(x) &=& \frac{\sqrt{|x|}}{2\sqrt{14}} + \frac{\nu}{2 |x|^{3/4}}\frac{q}{\sqrt{3}\cdot7^{1/4}} + \cdots  \quad (x \rightarrow -\infty) \nonumber
\end{eqnarray}

Upon integrating twice, the asymptotic expansions in equations
(\ref{0Bexpnsm=2}) and (\ref{0Bexpnsm=4}) furnish the free energy
perturbatively as an expansion in the dimensionless string coupling,
given by the same expression as before in
equation~\reef{0Astrcoupling}.

For these models, in the $\mu \rightarrow -\infty$ regime, $q$
represents the number of background ZZ D--branes in the model, with a
factor of $q$ for each boundary in the world sheet expansion, while in
the $\mu \rightarrow \infty$ regime it counts the number of units of
RR-flux in the background\cite{Klebanov:2003wg}. The asymptotic
expansions in the two directions can be argued in
ref.\cite{Klebanov:2003wg} to match onto each other analytically in a
particular ('t Hooft) limit. For the case $k=1$, the full
non--perturbative solution is known since it can be mapped directly to
the solution known for the $k=1$ type~0A case, as will be  discussed below
in  section~\ref{sec:connectingAB}.

For later reference, we briefly discuss the structure of these
solutions with increasing~$n$. As argued in ref.\cite{Klebanov:2003wg}
the $n=2$ expansions are deformations of the solutions of the
equation with $q=0$. However things get interesting for $n=4$. As before,
 the $x>0$ solution is a deformation of the solution with
$q=0$. For $x<0$, $q=0$ allows for the trivial solution $r(x)
= 0$, but trying to deform this for $q \neq 0$ leads only to a
complex solution. Additionally, $q=0$ allows for two nontrivial solutions
with $r(x) \neq 0$ and $\omega(x) \neq 0$. These two are related by
$\omega \rightarrow -\omega$ and are interpreted as $\IZ_2$ symmetry
breaking solutions. In the interpretation of
ref.\cite{Klebanov:2003wg}, this is due to the presence of R--R
fields. Both these solutions have a real extension to the case with $q
\neq 0$. The requirement of matching a $x<0$ solution to the $x>0$
solution picks out one of these for $q > 0$ and the other for $q <
0$. (The $x < 0$ solution listed above is for $q > 0$.) For higher $n$, more
such symmetry--breaking solutions arise. We will see how this is
organized explicitly in section~\ref{sec:beyond}.

\subsection{A Non--perturbative Connection Between 0A and 0B}
 \label{sec:connectingAB}

 It turns out that the simplest case of $n = 2$ ($k=1$), the 0A and 0B
 theories are non--perturbatively related in a very special way. In
 this case the conformal model is trivial ({\it i.e.}  ${\hat c}=0$)
 and we simply have the pure world--sheet supergravity sector. The
 strings are unencumbered by a spacetime embedding (not
 counting the ubiquitous Liouville direction,~$\phi$).

 The string equation for the $k=1$ 0A theory,
 equation~(\ref{streqn0A}) with $\mathcal{R} = w(z) - z$, is
\begin{equation}\label{streqn0Ak=1}
w\left(w-z \right)^2 - \frac{1}{2} \nu^2 \frac{\partial^2 w}{\partial z^2} \left(w-z\right) + \frac{1}{4} \nu^2 \left(\frac{\partial w}{\partial z} - 1\right)^2 = \nu^2 \Gamma^2 \quad .
\end{equation}
On the other hand, the string equation (\ref{streqn0B}) for the $k=1$
0B theory can be written succinctly as
\begin{equation}\label{streqn0Bm=2}
\nu^{2} \frac{\partial^2 r}{\partial x^2} -\frac{1}{2} r^3 + \frac{1}{2} x r + \nu^{2} \frac{q^2}{r^3} = 0 \quad . 
\end{equation}
Notice that the perturbative expansion for the $k=1$ 0A theory as $z
\rightarrow \infty$ (equation~\reef{0Aexpnsk=1}) looks similar to the
perturbative expansion for the $n=2$ 0B theory as $x \rightarrow
-\infty$ (equation~\reef{0Bexpnsm=2}) up to a (non--universal) sphere
term, once one identifies $\Gamma$ with $q$. The two expansions are
just offset by various powers of $2$.  In fact there exists a
non--perturbative map between the two
equations~\cite{Morris:1990bw,Morris:1992zr} that can be seen as
follows. First define a function $f(z)$ {\it via}
\begin{equation}\label{Morrismap}
w(z) = f(z)^{2} + z \quad ,
\end{equation}
for which the string equation for the 0A theory (\ref{streqn0Ak=1})
becomes
\begin{equation}\label{0Ak1to0Bm2}
\nu^2 \partial^{2}_{z} f - f^3 - z f + \nu^2 \frac{\Gamma^2}{f^3} = 0 \quad .
\end{equation}
After rescaling using
\begin{eqnarray*}
f = 2^{-1/6} r ; \qquad  z = 2^{-1/3} x \quad .
\end{eqnarray*}
equation (\ref{0Ak1to0Bm2}) becomes the string equation for the 0B
theory (\ref{streqn0Bm=2}), but with the sign of $x$ reversed. The
physics of pure 0A supergravity and pure 0B supergravity ({\it i.e.}
${\hat c}=0$) are non--perturbatively related, with the brane and flux
perturbative regions exchanged.

This non--perturbative connection between the 0A and 0B theories (for
$k=1$) follows, mathematically, from the fact that the same basic
string equation appears at the base of the two separate (KdV {\it vs}
ZS) hierarchies of equations. This connection is partly in the spirit
of a much larger set of connections that we are reporting on in this
paper. We find that the KdV and ZS structures are embedded in a much
larger structure, the dispersive water wave hierarchy of
equations, and find a class of connections (of a different sort)
between 0A and 0B for all~$k$ and see that they define two special
corners of a  larger tapestry of physical theories.

\section{The Dispersive Water Wave Hierarchy}
\label{sec:dww}

The standard dispersive water wave (DWW) hierarchy\cite{Kuper}, which
will play a central role in the new physics we uncover, is a
two--component system\footnote{See refs.\cite{Broer,Kaup1,Kaup2} for
  earlier studies of the properties of the dispersive water wave
  equations, and we will use the notation of
  refs.\cite{Kuper,Gordoa:2001}.}. It is described by:
  \begin{equation}\label{DWWPDE}
\mathbf u_{t_{n}}= R^{n}\mathbf u_{x} \equiv \nu\partial_x \mathbf{L}_{n+1}[\mathbf{u}] \quad ,
\end{equation}
where $\mathbf{u}_{t_n}\equiv \partial_{t_n}\mathbf{u}$, $\mathbf{u}_x \equiv \nu\partial_x\mathbf{u}$
(note that here and in the rest of the paper, for any function $G(x)$, $G_{x}$ will denote $\nu\,
\partial G/\partial x$), and we adopt a matrix notation:
\begin{eqnarray*}
\mathbf u=\left(\begin{array}{c}
                             u\\
                             v\end{array}\right) ,
\quad \quad
\mathbf{L}_n[\mathbf{u}]=\left(\begin{array}{c}
                             L_n[u,v]\\
                             K_n[u,v]\end{array}\right) \ .
\end{eqnarray*}
Here,
\begin{equation}\label{DWWRecOp}
R\equiv \frac{1}{2}\left(\begin{array}{cc}
                    \partial_{x}u\partial_{x}^{-1}-\partial_{x}&2\\
                    2v+v_{x}\partial_{x}^{-1}&u+\partial_{x}\end{array}\right) \quad ,
\end{equation}
is the recursion operator for the DWW hierarchy. The operator $R$ can
be written as the quotient of two Hamiltonian operators $B_{1}$ and
$B_{2}$,
\begin{eqnarray*}
R = B_{2}\circ B_{1}^{-1} \quad ,
\end{eqnarray*}
where $B_{1}$ and $B_{2}$ are given by
\begin{eqnarray} \label{formofBs}
B_{2}=\frac{1}{2}\left(\begin{array}{cc}
                    2\partial_{x}&\partial_{x}u-\partial_{x}^{2}\\
                    u\partial_{x}+\partial_{x}^{2}&v\partial_{x}+\partial_{x}v\end{array}\right)\ ,  \qquad
B_{1}=\left(\begin{array}{cc}
                    0&\partial_{x}\\
                    \partial_{x}&0\end{array}\right) \ .
\end{eqnarray}
The $\mathbf{L}_n$ obey the recursion relation
\begin{equation}\label{Lrec}
\mathbf{L}_{n+1,x}=R\,\mathbf{L}_{n,x}\, ,
\end{equation}
which follows immediately from (\ref{DWWPDE}). The first few $L_{n}$ and $K_{n}$ are as follows:
\begin{eqnarray}\label{DWWPolys}
L_0 &=& 2; \quad  K_0 = 0; \nonumber\\
L_1 &=& u; \quad  K_1 = v; \nonumber\\
L_2 &=& \frac{1}{2}u^2 + v - \frac{1}{2}u_x; \quad   K_2 = uv + \frac{1}{2}v_x;\nonumber\\
L_3 &=& \frac{1}{4} u^3 + \frac{3}{2} uv - \frac{3}{4} u u_x + \frac{1}{4} u_{xx};\\
K_3 &=& \frac{3}{4}u^2 v + \frac{3}{4}v^2 + \frac{3}{4} uv_x + \frac{1}{4}v_{xx} \quad .\nonumber
\end{eqnarray}
The normalization of $\mathbf{L}_0$ is chosen so as to reproduce (\ref{DWWPDE}) for $n=0$.\\
\indent The DWW hierarchy can be reduced to a one--component system by
demanding that one of the two independent functions vanish. If we set
$u(x)=0$, we actually reduce to the KdV hierarchy: two operations of
the (reduced) DWW recursion operator give
\begin{equation}\label{KdVRecOpfrmDWWRecOp}
  R\circ R\equiv R^{2}=\left(\begin{array}{cc}
      R_{2}&0\\
      \frac{1}{4}(2v_{x}+v_{xx}\partial_{x}^{-1})&R_{2}\end{array}\right) \quad ,
\end{equation}
where $R_{2}=\frac{1}{4}(\partial_{x}^{2}+4v+2v_{x}\partial_{x}^{-1})$
is the recursion operator of the KdV hierarchy, shown in
equation~\reef{KdVrec}. Thus we obtain a reduction of the even flows
of the original hierarchy (\ref{DWWPDE}) to
\begin{equation}\label{KdVPDE}
v_{t_{2n}}= R_{2}^{n}v_{x} \ ,
\end{equation}
which is the KdV system in $-v(x)$ with independent variable $x$, and
the even times of DWW map to the times of the KdV $t_{2n}\to
t_n$. (Compare  with equation~\reef{KdVflow}). \\
\indent In addition, the recursion relation~\reef{Lrec} reduces to
\begin{equation}\label{KdVrec2}
L_{2n+2,x}=R_2\circ L_{2n,x},
\end{equation}	
which is exactly the KdV recursion relation~\reef{KdVrec}. The relative normalizations
are $L_0=2$ and $P_0=\frac{1}{2}$ so we conclude that $L_{2n}=4P_k$. Moreover, with $u=0$, it
is immediate that $L_{2n+1}=0$.

The other obvious reduction, $v(x)=0$, reduces the system to
the Burgers hierarchy. We do not explore if there are any string
theory consequences of that in this paper, since $v(x)$ is used to
define the partition function of our theories in all our examples.

\subsection{Scaling Reductions and New String Equations}

Integrable hierarchies of partial differential equations (PDEs) can be reduced to ordinary differential equations
(ODEs) through an additional condition on the variables.  In our
context, these ODEs are sometimes to be thought of as defining string
theories; they are the ``string equations'' of a family of theories,
forming an hierarchy themselves. As outlined around
equation~\reef{KdVflow}, string equations can be thought of as
supplying the initial conditions for the partition function, and then
the PDEs of flows describe how the partition function evolves as a
function of the operators that couple to the $t_k$\cite{Douglas}.  The
original example of all
this\cite{Gross:1989vs,Brezin:1990rb,Douglas:1989ve} was a hierarchy
of equations that have the Painlev\'e~I equation as the non--trivial
equation at their base, indexed by an integer $k$. They defined the
bosonic $c<1$ string theories coupled to the $(2,2k-1)$ conformal
minimal models.

It was later realized\cite{Dalley:1991vr} that another rich family of
string equations (those in equations~\reef{streqn0A}) can be obtained
by imposing certain scaling relations on the variables of the KdV
system (note however that the equations were originally
derived\cite{Morris:1990bw,Dalley:1991qg} directly from matrix model
constructions analogous to the original route). With this in mind, we
explore a similarity reduction of the DWW hierarchy, expecting to
obtain new string equations at the end of the day.

We follow the approach originally used to derive the string equations
of type~0A~\reef{streqn0A} for the KdV hierarchy
\cite{Dalley:1991vr,Dalley:1992br}. To that end, assign $v$ mass
dimension 1. The dimensions of the other terms uniquely follow from
\reef{DWWPDE} and are $[u]=\frac{1}{2}$, $[x]=-\frac{1}{2}$ and
$[t_n]=-\frac{1}{2}(n+1)$ Thus we can write down two Callan--Symanzik
equations expressing the scaling symmetry,
\begin{equation}
\begin{split}
\frac{1}{2}u+\frac{1}{2}x u_x+\sum_{n=0}^{\infty}\frac{1}{2}(n+1)t_n u_{t_n} & = 0 \\
v+\frac{1}{2}x v_x+\sum_{n=0}^{\infty}\frac{1}{2}(n+1)t_n v_{t_n} & =0 \quad .
\end{split}
\end{equation}
Using~\reef{DWWPDE} and~\reef{Lrec} we can rewrite these equations,
\begin{equation}\label{CS1}
\begin{split}
\frac{1}{2}u+\frac{1}{2}x u_x+\sum_{n=0}^{\infty}\frac{1}{2}(n+1)t_n \left(\frac{1}{2}uL_{n,x}+K_{n,x}-\frac{1}{2}L_{n,xx}+\frac{1}{2}u_x L_n \right) & = 0 \\
v+\frac{1}{2}x v_x+\sum_{n=0}^{\infty}\frac{1}{2}(n+1)t_n\left(\frac{1}{2}v_x L_n+\frac{1}{2}K_{n,xx}+\frac{1}{2}u K_{n,x}+ v L_{n,x} \right) & =0\quad.
\end{split}
\end{equation}
Defining,
\begin{equation}\label{LK}
\left(\begin{array}{c}
                             \mathcal{L}\\
                             \mathcal{K}\end{array}\right) = \sum_{n=0}^{\infty}\frac{1}{2}(n+1)t_n \mathbf{L}_n,
\end{equation}
we can rewrite~\reef{CS1},
\begin{eqnarray}
\label{CS21}
\frac{1}{2}u\mathcal{L}_x+\frac{1}{2}u_x \mathcal{L}+\mathcal{K}_x-\frac{1}{2}\mathcal{L}_{xx}&=& 0 \\
\label{CS22}
v\mathcal{L}_x+\frac{1}{2}v_x \mathcal{L}+\frac{1}{2}\mathcal{K}_{xx}+\frac{1}{2}u\mathcal{K}_x&=&0\quad,
\end{eqnarray}
where we have used the fact that we will take $t_0=x$ and the other
$t_n$ to be independent of $x$.  Equation~\reef{CS21} can readily be
integrated. Moreover, solving~\reef{CS21} for $\mathcal{K}_x$ and
substituting the result into~\reef{CS22} yields an expression, which,
after multiplying by $\mathcal{L}$, can also be integrated. The
results are our new coupled string equations,
\begin{eqnarray}\label{DWWstring1}
-\frac{1}{2}\mathcal{L}_x+\frac{1}{2}u\mathcal{L}+\mathcal{K} &=& \nu c\quad \quad \quad \quad \quad  \\
\label{DWWstring2}
\left(-v+\frac{1}{4}u^2+\frac{1}{2}u_x\right)\mathcal{L}^2-\frac{1}{2}\mathcal{L}\mathcal{L}_{xx}+\frac{1}{4}\mathcal{L}_x^2&=&\nu^2 \Gamma^2\quad ,\quad \quad \quad \quad \quad
\end{eqnarray}
where we have introduced two integration constants, $c$ and
$\Gamma$. We stress that the simplest possibility is for $c$ and
$\Gamma$ to be independent of $x$ \emph{and} $t_i$, though only
independence of $x$ is strictly necessary.
\\
\indent The $n$th model is chosen by setting all $t_i$ equal to zero
except for $t_0=x$ and $t_n$ which is chosen to be a numerical factor
to fix the normalization. We choose to parameterize $t_n$ as
\begin{equation}\label{tntogn}
 g_n \equiv \frac{1}{\frac{1}{2}(n+1)t_n}
\end{equation}
in order to make direct contact with some recent literature which
discusses this system in a much different (mathematical)
context\cite{Gordoa:2001}.

\section{The Organizing Role of Painlev\'e IV}
\label{sec:painleve}
Let us focus on the case $n=1$, which forms the bottom of the
hierarchy of string equations from which all others follow using the
recursion relations. The string equations in this
case reduce to the Painlev\'e IV equation, an important equation from
the mathematical literature. Its appearance at the bottom of the
ladder of string equations we're presenting here is significant. Note
that the entire family of string equations can be generated from this
$n=1$ case by use of the recursion operator, and so structures at this
level will be reflected at higher $n$, even while the complexity of
the equations increases. Also notable is that this is the first time
that a role for this equation has been uncovered in this context of
non--perturbative string theory, and it takes its place alongside the
Painlev\'e~I and~II equations whose roles (mentioned earlier) have
been established in this context already. In fact, part of the
motivation that led to the discoveries upon which we report here was
the question as to the further role of the Painlev\'e equations in
such systems. Painlev\'e~IV emerged naturally as a candidate equation
to play a role and this led to our studying of the DWW system that we
found connected to Painlev\'e~IV in the literature\cite{Gordoa:2001}.

Let us see more explicitly how the equation emerges. Remarkably, it
will naturally appear in two different ways\cite{Gordoa:2001,Gordoa:2005}.

\subsection{Painlev\'e~IV: First Movement}
The string equations for $n=1$ are:
\begin{eqnarray}\label{DWWn=1}
2v - u_x + u^2 + g_1 x u = 2 \nu g_1 (c + \frac{1}{2}) \quad , \hspace{30mm} \\
\left(-v + \frac{1}{4}u^2 + \frac{1}{2}u_x\right)(u + g_1 x)^2 -\frac{1}{2}u_{xx}(u + g_1 x) + \frac{1}{4}\left(u_x + \nu g_1\right)^2  = \nu^2 g_1^2 \Gamma^2 \quad . \nonumber
\end{eqnarray}
where we have used the relation $g_1 = \frac{1}{t_1}$.

Solving the first of these for $v$ gives
\begin{equation}\label{DWWn=1v}
v = \frac{1}{2}\left(u_x - u^2 - g_1 x u + 2 \nu g_1 (c + \frac{1}{2})\right) \quad ,
\end{equation}
and substituting this into the second yields a second order ODE in $u$
which, under the change of variables
\begin{equation}
u(x) = y(x) - g_1 x \quad
\end{equation}
becomes
\begin{equation}
y_{xx} = \frac{1}{2}\frac{y_{x}^2}{y} + \frac{3}{2}y^3 -2 g_1 x y^2 + 2 \left[\left(\frac{g_1^2x^2}{4}\right) - \nu \alpha_1\right] y - \nu^2 \frac{1}{2} \frac{\beta^2_1}{y} \quad .
\end{equation}
$\alpha_1$ and $\beta_1$ are constants related to $c$ and $\Gamma$ in
the DWW string equations through
\begin{eqnarray}\label{alphabetaDefn}
\alpha_{1} = g_{1}(c + \frac{1}{2})\ , \qquad
\beta_{1} = \pm 2 g_{1} \Gamma \ ,
\end{eqnarray}

Setting $g_1 = -2$, and dropping the subscripts on the constants, gives
\begin{eqnarray}
\label{eq:painleve}
  y_{xx} = \frac{1}{2}\frac{y_{x}^2}{y} + \frac{3}{2}y^3 +4 x y^2 + 2 \left(x^2 - \nu \alpha\right) y - \nu^2 \frac{1}{2} \frac{\beta^2}{y} \ ,
\end{eqnarray}
which is the fourth Painlev\'e equation $P_{IV}$ in standard form, and
\begin{equation}
  \alpha=-(2c+1)\ ,\quad \beta=-8\Gamma^2\ . \label{eq:painleveconstants1}
\end{equation}

We will see in the next section, specific constraints yielding the 0A and 0B
theories that require $c = -\frac{1}{2} $ and $\Gamma = 0$ respectively. Notably,
these are precisely the values for which the constants $\alpha$ and
$\beta$ in the standard form of Painlev\'e~IV vanish.

\subsection{Painlev\'e~IV: Second Movement}

In fact, there is another natural appearance of the Painlev\'e~IV
equation in this system\cite{Gordoa:2001,Gordoa:2005}, at $n=1$. There is a
natural generalization\cite{Kuper} of the Miura map (that we saw for
KdV in section~\ref{sec:type0}) to the DWW system, defining
new variables $U$ and $V$:
\begin{equation}
  u=U\ , \qquad v=UV-V^2+V'\ .
\end{equation}
Now, as we saw above, $y(x)= u(x)-2x = U(x)-2x$ satisfies
Painlev\'e~IV with constants $\alpha$ and $\beta$ given in
equation~\reef{eq:painleve}. Well, additionally, $-V(x)$ satisfies a
copy of Painlev\'e~IV (equation~\reef{eq:painleve}) also, but with
constants related to our physical parameters by
\begin{equation}
\label{eq:painleveconstants2}
\alpha=\mp 3\Gamma+c+1\ ,\quad \beta=(c\pm\Gamma)^2\ .
\end{equation}

We take this seriously, not the least because the variables described
by the Miura map in the case of KdV (type~0A) were seen to be
physically very natural, pertaining as they do to the FZZT and ZZ
D--branes. (See the end of section~\ref{sec:0A} for a brief review.)
We expect therefore (but this needs more exploration) that this DWW
Miura map also leads to rich physics. The cases $c=\pm\Gamma$ and
$c=-1\pm3\Gamma$ imply vanishing of $\alpha$ and $\beta$ and may well
have some special significance in this context. (We will, for example,
find special solutions for all $n$ corresponding to  $c=\pm\Gamma$
points. It is also interesting to note that the equations together
point to the values $c=\pm\frac12$, $\Gamma=\pm\frac12$, values which
do feature prominently in what is to follow.)

\section{Connecting the Type~0 String Theories}
\label{sec:connectAB}

Having introduced the DWW hierarchy, we now show how {both} the
type~0A and type~0B string theories coupled to the $(2,4k)$
superconformal minimal models can be found embedded in this system. We
show that by placing appropriate constraints on the full system of
string equations, one can recover the respective string equations for
both of the type~0 theories. Quite beautifully, these constraints require one of the
two integration constants $(c,\Gamma)$ to freeze to particular
values. The remaining unfixed constant then acts as the parameter that
counts the number of ZZ--branes or units of R--R flux in each theory,
depending on which asymptotic region (positive or negative large $x$)
is under consideration.

\subsection{Reduction to 0A}
It was seen in section~\reef{sec:dww} that setting $u$=$0$ reduces the DWW
hierarchy to the KdV hierarchy. We therefore expect that this constraint also
reduces our new string equations to the~0A string equations. That this indeed
occurs can be seen as follows. Equation~\reef{LK} gives,
\begin{equation}
\begin{split}
\mathcal{L}[u\mathrm{=}0,v\mathrm{=}-w]&=\sum_{n=0}^{\infty}\frac{1}{2}(n+1)t_n^{\mathrm{DWW}}L_n \\
&=\sum_{n=0}^{\infty}\frac{1}{2}(2n+1)t_{2n}^{\mathrm{DWW}}L_{2n} \\
&=4\sum_{j=0}^{\infty}(j+\frac{1}{2})t_{2j}^{\mathrm{DWW}}P_{j} \\
&=\mathcal{R}[w]
\end{split}
\end{equation}
where we have used that $L_{2n+1}[u$=$0,v]=0$ and $L_{2n}[u$=$0,v$=$-w]=4P_n[w]$ (see below~\reef{KdVrec2}). The last equality holds provided that we make the identification,
\begin{equation}\label{tn}
t_{2n}^{DWW}= \frac{1}{4}t_n^{\mathrm{KdV}}=\frac{(-1)^{n+1}4^n(n!)^2}{(2n+1)!}
\; \Rightarrow \;g_{2n} = 2\frac{(-1)^{n+1}(2n)!}{4^n(n!)^2}\quad .
\end{equation}
Finally, we see that when $u=0$ and $v=-w$, equation~\reef{DWWstring2}
exactly reduces to equation~\reef{streqn0A}, \emph{i.e.}  our new
string equations encode 0A string theory coupled to the $(2,4k)$
superconformal minimal models. For even flows it is easy to show that
$u=0$ is only consistent with the other string
equation~\reef{DWWstring1} if $c$ is frozen:
\begin{equation}
c = -\frac{1}{2} \ .
\end{equation}
So one of the parameters in the original DWW equations becomes fixed
when recovering type~0A coupled to the $(2,4k)$ superconformal minimal
models, while the other parameter~$\Gamma$ counts the number of branes
or units of RR--flux in the type 0A theory. We will see this behaviour
again in the case that we recover the type~0B theory.

\subsection{Reduction to 0B}\label{sec:0Breduc}
Consider the following redefinition of the DWW variables $\{u(x), v(x), x\}$ to the ZS
variables $\{r(y), \omega(y), y\}$:	
\begin{eqnarray}\label{DWWto0B}
y=2x,\quad
u(y)=2\left(\omega(y) - \frac{r_{y}}{r(y)}\right), \quad
v(y)= -r^2(y) \  .
\end{eqnarray}
The recursion relation~\reef{Lrec} becomes,
\begin{equation}
\left(\begin{array}{c}
                             L_{n+1}\\
                             K_{n+1,y}\end{array}\right) =
\left(\begin{array}{c}
                             (\omega-\frac{r_y}{r})L_n-L_{n,y}+K_{n,y}\\
                             -r(rL_n-\frac{K_{n,y}}{r})_y+\omega K_{n,y}\end{array}\right),
\end{equation}
or,
\begin{equation}
\left(\begin{array}{c}
                             R_{n+1}\\
                             H_{n+1,y}\end{array}\right) =
\left(\begin{array}{c}
                             \omega R_n-\left(\frac{H_{n,y}}{r}\right)_y+rH_{n,y}\\
                             -r R_{n,y}+\omega H_{n,y}\end{array}\right),
\end{equation}
where we have defined,
\begin{equation}
R_n =\frac{1}{2}\left( r L_n-\frac{K_{n,y}}{r}\right),\quad\quad H_n =\frac{1}{2}K_n.
\end{equation}
These are precisely the recursion relations of the ZS
hierarchy~\reef{ZSrec}. Moreover, the $H_n$ and $R_n$ just defined
actually agree with those presented in~\reef{ZSpoly}. It suffices to
check $n=0$: from~\reef{DWWPolys} we have $L_0=2$ and $K_0=0$ which
imply $H_0=0$ and $R_0=r$, as expected.
\\
\indent Finally, we may ask how we can produce the~0B string
equations~\reef{streqn0B} from our new string
equations~\reef{DWWstring1} and~\reef{DWWstring2}. The answer turns
out to be simple and elegant: all we must do is set
\begin{equation}
\mathcal{L}=0\quad .
\end{equation}
Equation~\reef{DWWstring1} then requires,
\begin{equation}
\sum_{n=0}^{\infty}t_n(n+1)H_n-\nu c=0\quad,
\end{equation}
which further implies,
\begin{equation}
 \sum_{n=0}^{\infty}t_n(n+1)R_n = 0\quad .
\end{equation}
The $t_n$ required here to consistently produce the equations of \cite{Klebanov:2003wg} are identical to the values we determined earlier~\reef{tn}. So, upon identifying $c=-q$, we have exactly produced the~0B string equations. \\
\indent Again notice how the consistency of our constraint
$\mathcal{L}=0$ with~\reef{DWWstring2} forces one of our parameters to
vanish
\begin{equation}
\Gamma = 0 \quad ,
\end{equation}
leaving the parameter $c = -q$ to count the number of ZZ branes or
RR--fluxes in the type 0B theory. Finally, we remark that the partition function
of the~0B theory is determined \emph{via}
\begin{equation}
\begin{split}
F &= \frac{1}{\nu^2}\int d^2y\; \frac{r(y)^2}{4}\\
&=-\frac{1}{\nu^2}\int d^2x\; v(x)
\end{split}
\end{equation}
so that $-v(x)$ encodes the partition function for both~0A and~0B.

\section{DWW Unconstrained --- Beyond the Familiar}
\label{sec:beyond}
We have seen that constraining the DWW string
equations~\reef{DWWstring1} appropriately leads to the 0A and 0B
theories (coupled to the superconformal $(2,4k)$ series),
respectively. The constraints take the system with two free parameters
$(c,\Gamma)$ and define two special points: 0A with $(c=-1/2, \Gamma$ free)
and 0B with $(c$ free, $\Gamma{=}0)$. We can also consider the fully
unconstrained system with both parameters $(c,\Gamma)$ unfixed, and
general $\{v(x), u(x)\}$. Interestingly, we get {\it multiple}
asymptotic expansions for the variable $v(x)$. The structure of the
equations and the corresponding asymptotic expansions gets richer as
$n$ increases. Since in both cases, asymptotic expansions of $v(x)$
gave us, upon integrating twice, an expansion of a partition function
for a string theory, we look again for it to define an interesting
partition function in the new cases we will encounter. While this is
an assumption, we shall see it bear fruit presently.

In this section we will first list the asymptotic expansions for the
first few $n$ obtained from the corresponding string equations. We
will explain the organizational rules we use to group these
expansions into various classes.  A careful analysis of the patterns
we uncover in what follows allows us to extrapolate to higher $n$ and
predict the structure of the expansions for any $n$.  We will see that a
subset of these, when appropriately combined, reproduce the type~0
expansions that we have already encountered. In addition, we obtain
{\it completely new} expansions which have not been presented in the
literature before. Our key observation here is that these also
resemble perturbative sectors of string theories (either with branes
or fluxes present). We take these seriously as new string theories and
our task after this section will be to identify what string theories
they might be.

The number of expansions grows large as $n$ increases (we will see
later that the number of expansions is $(n+1)^2$). To deal with this
proliferation of expansions, we classify them into classes whose
members are related to one another by simple symmetries. The classes
themselves are distinguished by a number of salient features, many of
which we explore in what follows. We choose to define the classes
based on their behavior at order $\nu^0$ (this is equivalent to the
leading behavior in $g_s^{-2}$, the sphere level of closed string
perturbation theory, as we will see later). Since DWW is a two
component system, we must consider the leading behavior of both
functions, $u$ and $v$. We adopt the following
classification scheme:\\
\begin{equation}\label{classes}
\begin{split}
\textrm{Class 1:}\quad&u_1 \sim 0,\quad\quad\; v_1\sim x^{2/n}\\
\textrm{Class 2:}\quad&u_2 \sim 0,\quad\quad\; v_2\sim 0\\
\textrm{Class 3:}\quad&u_3 \sim x^{1/n},\quad v_3\sim 0\\
\textrm{Class 4:}\quad&u_4 \sim x^{1/n},\quad v_4\sim x^{2/n},\quad u_4^2/v_4 \sim 1/4\\
\textrm{Class 5:}\quad&u_5 \sim x^{1/n},\quad v_5\sim x^{2/n},\quad u_5^2/v_5 \sim a\neq 1/4\\
\end{split}
\end{equation}
\\We postpone the detailed study of $u$ to subsequent work. Here we
mention its leading behavior only to complete the classification; in
what follows, we focus exclusively on $v$ and its asymptotic
expansions.

The details of sections~\ref{sec:expandone}--\ref{sec:expandfour},
being a list of examples that we found instructive, might be a little
dry on first reading and so the reader is encouraged to skip to
section~\ref{sec:general} for the general case.

\subsection{$n = 1$}
\label{sec:expandone}
The string equations for this case were already written in
equations~\reef{DWWn=1}.  Solving the first of these for $v$ gave
equation~\reef{DWWn=1v}, and substituting into the second yields a
scalar second order ODE in $u(x)$ (equivalent to Painlev\'e~IV), which
can be used to produce the expansions. Asymptotic expansions for $u(x)$
can then be used to yield asymptotic expansions for $v(x)$ using
equation~\reef{DWWn=1v}.

We obtain three classes of expansions for $v(x)$, 
\begin{eqnarray}\label{DWWn=1expn}
v_2&=&\frac{\nu^2}{x^2}(c^2-\Gamma^2)\left(1-\frac{\nu}{ g_1 x^2}6c+\frac{\nu^2}{g_1^2 x^4}(45c^2-5\Gamma^2+5)-\cdots\right) \quad ,\nonumber\\
v_3&=&\nu(c-\Gamma)\left(1-\frac{\nu}{g_1 x^2}2\Gamma-\frac{\nu^2}{g_1^2 x^4}6\Gamma(c-3\Gamma)-\cdots\right) \quad ,\\
v_4&=&\frac{1}{9}g_1^2 x^2+\nu\frac{2 g_1 c}{3}-\frac{\nu^2}{x^2}\frac{1}{3}(3c^2+9\Gamma^2-1)+\frac{\nu^3}{g_1
x^4}6c(c^2-9\Gamma^2) - \cdots \quad .\nonumber
\end{eqnarray}

Upon integrating twice (following what we learned from the type~0
theories in sections~\ref{sec:0A} and~\ref{sec:0B}), one can obtain the
free energy for a genus expansion of a
string theory which allows us to identify the string coupling to be
$g_{s} = {\nu}/{x^2}$.

\subsubsection{Symmetries for $n=1$}
The other expansions within each class can be obtained by the following
symmetry operation:
\begin{eqnarray*}
f_1: \Gamma \to -\Gamma \quad .
\end{eqnarray*}
Since $v_2$ and $v_4$ contain only even powers of $\Gamma$, the are
invariant under this map; however, $f_1\circ~v_3~\neq~v_3$. Thus there are
two expansions in the $v_3$ class, and, together with $v_2$ and $v_4$,
these comprise {four} total $n=1$ expansions.

\subsection{$n = 2$}
The string equations are:
\begin{eqnarray}\label{DWWn=2}
u_{xx} &=& 3u u_x - u^3 -6u v -2g_2 x u + 4 \nu g_2\left(c + \frac{1}{2}\right) \quad ,\nonumber\\
v_{xx} &=& 2\left( \frac {(u v + \frac{1}{2}v_x - \nu g_2 c)^2 - \nu^2 {g_2}^2 {\Gamma}^2}{v+ \frac{1}{2}u^2 -\frac{1}{2}u_x + g_2 x}\right) \\
&&- 2v \left(v+\frac{1}{2}u^2-\frac{1}{2}u_x + g_2 x\right)- 2(u v)_x \quad . \nonumber
\end{eqnarray}
where again $g_2 = \frac{1}{t_2}$. Solving the first of these for $v$ gives,
\begin{equation}\label{DWWn=2v}
v = \frac{1}{6 u}\left(u_{xx} -3u u_x + u^3 + 2g_2 x u -4 \nu g_2 \left(c + \frac{1}{2}\right)\right) \quad .
\end{equation}
Substituting this into the second yields a scalar fourth order ODE in
$u$, which can then be used to produce the expansions for $v$. In this
case, there are four relevant classes of expansions
\begin{eqnarray}\label{DWWn=2expn}
v_1&=&-g_2x-\frac{\nu g_2^{1/2}}{x^{1/2}}\Gamma+\frac{\nu^2}{x^2}\frac{1}{8}\left(-4c^2+4\Gamma^2+1\right) + \cdots \quad ,\nonumber\\
v_2&=&\frac{\nu^2}{x^2}\left(c^2-\Gamma^2\right)\left(1-\frac{2\nu^2}{g_2 x^3}(5c^2-\Gamma^2+1)+ \cdots \right) \quad ,\\
v_3&=&\frac{g_2^{1/2}\nu}{x^{1/2}}(c-\Gamma)\left(\frac{i}{\sqrt{2}}+\frac{\nu}{g_2^{1/2}x^{3/2}}\frac{1}{4}(c-5\Gamma)- \cdots\right) \quad ,\nonumber\\
v_4&=&-\frac{g_2x}{5}+\frac{\nu g_2^{1/2}}{x^{1/2}}\frac{ic}{\sqrt{5}}-\frac{\nu^2}{ x^2}\frac{1}{4}(2c^2+10\Gamma^2-1) -\cdots \quad .\nonumber
\end{eqnarray}
Here, we see that the string coupling is $g_{s} = {\nu}/{x^{\frac{3}{2}}}$.

\subsubsection{Symmetries for $n=2$} The other expansions within each class can
be obtained by the following operations,
\begin{eqnarray*}
f_1:\Gamma\to-\Gamma \ ,\qquad
f_2:c\to-c \  ,
\end{eqnarray*}
and compositions thereof.  Some of the $v_i$ are invariant under one
or both of these maps.
Altogether, there are {nine} distinct expansions.  Here are the
four classes of expansions together with the number of distinct
expansions within each class and the maps that lead to them:
\begin{eqnarray*}
v_1(2) : \{1, f_1\}; \quad v_2(1): \{1\}; \quad v_3(4):\{1, f_1, f_2, f_1\circ f_2\}; \quad v_4(2):\{1, f_2\} \quad .
\end{eqnarray*}

\subsection{$n=3$}

The string equations are too long to be written down explicitly here
and so we omit them.  Four classes of expansions are produced in this
case, and one sees expansions in Class 5 appearing for the first time here.
\begin{eqnarray}
v_2 &=&\frac{\nu^2}{x^2}(c^2-\Gamma^2)\left(1-\frac{\nu^3}{g_3 x^4}\frac{5}{2}c(7c^2-3\Gamma^2+5) + \cdots \right) \quad , \nonumber\\
v_3&=&\frac{\nu}{x^{2/3}}(c-\Gamma)\left(\frac{(-2)^{2/3} g_3^{1/3}}{3}+\frac{\nu}{x^{4/3}}\frac{1}{3}(c-3\Gamma)- \cdots\right) \quad , \\
v_4&=&\frac{2\cdot 2^{1/3}}{35^{2/3}}g_3^{2/3}x^{2/3}+\frac{\nu g_3^{1/3}} {x^{2/3}}\frac{2\cdot2^{2/3} c}{3\cdot35^{1/3}} - \frac{\nu^2}{ x^2}\frac{1}{9}(3c^2+21\Gamma^2-2) + \cdots \quad , \nonumber\\
v_5&=&-\frac{2\cdot2^{1/3}}{5^{2/3}}g_3^{2/3}x^{2/3}-\frac{\nu g_3^{1/3}}{x^{2/3}}\frac{2^{2/3}}{3\cdot5^{1/3}}(c+\sqrt{5}\Gamma)-\frac{\nu^2}{x^2}\frac{1}{9}(3c^2-3\Gamma^2-1) + \cdots \quad .\nonumber
\end{eqnarray}
Here the string coupling turns out to be  $g_{s} = {\nu}/{x^{\frac{4}{3}}}.$

\subsubsection{Symmetries for $n=3$}
The other expansions within each class can
be obtained by the following operations,
\begin{equation}
f_1:\Gamma\to-\Gamma \ , \quad f_3:x\to-x \ , \quad f_4:g_3\to-g_3\ ,
\end{equation}
and any compositions of these maps.
A quick calculation shows that there 16 different expansions
\begin{eqnarray*}
v_2(1)&:&\{1\} ; \nonumber\\
v_3(6)&:& \{1, f_1, f_3=f_4, f_1\circ f_3 = f_1\circ f_4, f_1\circ f_3 \circ f_4, f_3 \circ f_4\}; \nonumber\\
v_4(3)&:& \{1, f_3=f_4, f_3\circ f_4\}; \nonumber\\
v_5(6)&:& \{1, f_1, f_3=f_4, f_1\circ f_3 = f_1\circ f_4, f_1\circ f_3 \circ f_4, f_3 \circ f_4\} \quad .\nonumber
\end{eqnarray*}

\subsection{$n = 4$}
\label{sec:expandfour}
Our explicit string equations are rather complicated and so we will
not list them here. The following five classes of expansions are
obtained:
\begin{eqnarray}\label{DWWn=4expn}
v_1(x) &=& -\frac{2 i }{\sqrt{3}}\sqrt{g_4}\sqrt{x} - \frac{\nu g_4^{1/4}}{x^{3/4}} \frac{(1+i)\Gamma }{2 \cdot 3^{1/4}} -\frac{\nu^2}{x^2}\frac{1}{48}\left(12 c^2-12\Gamma^2-5\right) + \cdots \quad ,\nonumber \\
v_2(x) &=&-\frac{\nu^2}{x^2}(\Gamma^2-c^2)\left(1-\frac{3\nu^4}{2 g_4 x^5}(21c^4-14 c^2 \Gamma^2 + \Gamma^4 + 35 c^2 - 5\Gamma^2 + 4)\right) + \cdots \quad ,\nonumber\\
v_3(x) &=& -\frac{g_4^{1/4} \nu}{x^{3/4}}(c-\Gamma)\left(\frac{1}{2^{3/4}(1+i)}+\frac{\nu}{g_4^{1/4} x^{5/4}}\frac{7\Gamma-3c}{8} + \cdots \right) \quad ,\\
v_4(x)&=&-\frac{2i}{3\sqrt{7}}\sqrt{g_4}\sqrt{x}-\frac{g_4^{1/4}\nu}{x^{3/4}}\frac{c}{\sqrt{3}\cdot7^{1/4}(1+i)}-\frac{\nu^2}{ x^2}\frac{1}{24}\left(6c^2+54\Gamma^2-5\right) + \cdots \quad ,\nonumber\\
v_5(x)&=&-2\sqrt{\frac{2}{21}}\sqrt{g_4}\sqrt{x}-\frac{g_4^{1/4}\nu}{x^{3/4}}\frac{21^{1/4}}{2^{5/4}}\left(-\frac{c}{\sqrt{7}}+\frac{\Gamma}{\sqrt{3}}\right)-\frac{\nu^2}{ x^2}\frac{1}{48}\left(12c^2-12\Gamma^2-5\right) + \cdots\nonumber \quad .
\end{eqnarray}
Here, the  string coupling is $g_{s} = {\nu}/{x^{\frac{5}{4}}}$.

\subsubsection{Symmetries for $n=4$}
The other expansions within each class can
be obtained by the following operations,
\begin{equation}
f_1: \Gamma \to -\Gamma \ , \quad f_2:c \to -c \ , \quad f_{3,4}: (x, g_4) \to (-x, -g_4) \ ,
\end{equation}
and any arbitrary composition of those maps.
A quick calculation shows that there are 25 distinct
expansions.  Here are the five classes of expansions together with the
number of distinct expansions within each class and the maps that lead
to them:
\begin{eqnarray*}
v_1(4) &:& \{1, f_1, f_{3,4}, f_1 \circ f_{3,4}\}; \nonumber\\
\quad v_2(1) &:& \{1\}; \nonumber\\
\quad v_3(8) &:& \{1, f_1, f_2, f_1\circ f_2, f_{3,4}, f_1\circ f_{3,4}, f_2 \circ f_{3,4}, f_1\circ f_2 \circ f_{3,4}\}; \nonumber\\
v_4(4) &:& \{1, f_2, f_{3,4}, f_2 \circ f_{3.4}\}; \nonumber\\
\quad v_5(8) &:& \{1, f_1, f_2, f_1\circ f_2, f_{3,4}, f_1\circ f_{3,4}, f_2 \circ f_{3,4}, f_1\circ f_2 \circ f_{3,4}\} \quad . \nonumber
\end{eqnarray*}

\subsection{Patterns  and Asymptotia}
\label{sec:general}
The expansions displayed above for $n=1$ to $n=4$ exhibit a rich
structure which we explore shortly. First we briefly review the
interpretation given to the parameter $\Gamma$ of the~0A theory (and
also to $q$ of the~0B theory). In the $\mu$ (or $x$) $\rightarrow
+\infty$ regime, $\Gamma$ represents the number of background ZZ
D--branes in the model, with a factor of $\Gamma$ for each boundary in
the worldsheet expansion. Since an orientable surface with odd (even)
Euler characteristic must contain an odd (even) number of boundaries,
$\Gamma$ must be raised to an odd (even) power if $g_s$ is. In
addition, the power of $\Gamma$ must be less than or equal to the
power of $g_s$. On the other hand, in the $\mu$ (or $x$)~$\rightarrow
-\infty$ regime, $\Gamma$ represents the number of units of RR--flux
in the background, with $g_s^2 \Gamma^2$ appearing when there is an
insertion of pure RR--flux. So in this case both $\Gamma$ and~$g_s$
should appear with even powers.
\\
\indent In applying these observations to our DWW expansions, we
immediately notice the remarkable fact that the various expansions
have powers of the parameters which somehow allow for interpretations
as counting branes or fluxes.  This is by no means guaranteed, and
indeed its occurrence was one of our main motivations for in--depth
study of the system. The presence of two parameters, however, leads to
a few subtleties. For example, in some expansions an interpretation in
terms of branes is only possible if one of the two parameters is set
to zero. With keep such observations in mind as we begin the study of
the various expansions.

Finally we note that the asymptotic direction ({\it i.e.,} positive or
negative $x$) of each expansion can be fixed by requiring that once we
fix the value of $g_n$, the expansion must be real (which is an
important constraint since $v$ encodes the free energy).  The value of
$g_n$, in turn, can be fixed using the values listed in
equation~(\ref{tn}) since we must reproduce the 0A theory. With all of
these observations we are ready to begin analyzing our expansions.

\subsubsection{Class 1}
\begin{itemize}
\item $v_1$ contains powers of $\Gamma$ consistent with those of  a parameter
  counting branes. This remains true for any value of $c$.
\item $v_1$ contains powers of $c$ consistent with those of  a parameter
  counting fluxes. However, for arbitrary $\Gamma$, $g_s$ appears with
  odd powers, inconsistent with  our requirements for a description of
  fluxes, as mentioned above. This problem is avoided if we set
  $\Gamma=0$ since this forces the odd powers of $g_s$ to vanish.
\item In particular, setting $\Gamma = 0$ with $g_n$ given by
  equation~\reef{tn} reduces this expansion to the $x>0$
  flux-expansion in $c$ (or $q$) for the type~0B theory seen in
  equations~(\ref{0Bexpnsm=2},\ref{0Bexpnsm=4}) for $n = 2,4$
  respectively.
\item Alternatively, setting $c = -\frac{1}{2}$ with the same value of
  $g_n$ reduces these expansions to those of the type~0A for
  $x>0$. (We listed them in equations~(\ref{0Aexpnsk=1} and
  \ref{0Aexpnsk=2}) for $k=1,2$ respectively\footnote{Recall that the
    DWW hierarchy index $n$ is related to KdV hierarchy index $k$ by
    $n=2k$.}.)
\item With the values of $g_n$ need to reduce to~0A and~0B, one
  obtains real expansions in this class {\it only} if $x > 0$. Hence
  we fix this class of expansions to be $x \rightarrow +\infty$
  asymptotic expansions.
  \item This class of expansions only exists for even $n$.
\end{itemize}
\subsubsection{Class 2}
\begin{itemize}
\item For even $n$, $v_2$ contains powers of $\Gamma$ and $g_s$
  consistent with those of a parameter counting fluxes. This is true
  for all values of $c$. For odd $n$, the powers of $\Gamma$ are still
  consistent with the flux interpretation, but there are odd powers of
  $g_s$ which are inconsistent with fluxes. These odd powers can be
  removed by setting $c=0$.
\item For even $n$, $v_2$ contains powers of $c$ and $g_s$ consistent
  with a parameter counting fluxes. This is true for all values of
  $\Gamma$. For odd $n$, the powers of $c$ are consistent with those
  of a parameter counting branes. In this interpretation, there are no
  contributions from surfaces with only one boundary.
\item Setting $c =-\frac{1}{2}$ with $g_n$ as chosen in
  equation~\reef{tn}, reduces the $v_2$ expansions to the type~0A
  expansions for $x<0$. (We listed them in equations~\reef{0Aexpnsk=1}
  and \reef{0Aexpnsk=2} for $k=1,2$ respectively.)
 \item Only even powers of $g_n$ and $x$ appear, so the
  requirement of reality does not fix the direction of these
  expansions.
  \item Consistency with the 0A expansions forces us to consider
  the expansions in this class as $x \rightarrow -\infty$
  expansions. We note the possibility
  that these might appear as $x \rightarrow +\infty$ expansions
  outside of the simple type~0A context we've seen so far\footnote{In
    fact, we can already think of an example. There are rational
    solutions of the type~0A string equations that were considered in
    a string theory context in ref.\cite{Johnson:2006ux}. The rational
    solutions have $v_2$ type expansions (for $c=-1/2$) in both
    asymptotic directions for $x$. Clearly there are analogous
    rational solutions for the full DWW equations that have $v_2$
    asymptotic expansions that generalize the known cases. We have
    constructed large families of them, and leave their study for a
    later publication.}.
\item These expansions vanish when $c^2
  = \Gamma^2$. (This is a likely special point(s) in parameter space. We got a
  first hint of this point in section~\ref{sec:painleve} where the
  second copy of Painlev\'e~IV has $\beta=0$.)
    \item This class of expansions appears for all $n$.
\end{itemize}
\subsubsection{Class 3}
\begin{itemize}
\item $v_3$ contains powers of $\Gamma$ consistent with those of a
  parameter counting branes. This is true only for $c=0$.
\item $v_3$ contains powers of $c$ consistent with those of a
  parameter counting branes. This is true only for $\Gamma=0$.
\item We notice that associating one boundary to each factor of $c$
  \emph{and} $\Gamma$ also produces a consistent worldsheet
  expansion. We might speculate about whether, in general, these
  expansions might capture $c$ and $\Gamma$ simultaneously counting
  branes.
\item Setting $\Gamma = 0$ with $g_2 = 1$ reduces these expansions to
  the $x<0$ brane-expansions for the type~0B theory for $n=2$ as seen
  in equation~(\ref{0Bexpnsm=2}). Hence we fix the expansions in this
  class to be $x \rightarrow -\infty$ expansions.
\item At $n=4$, the value $g_4 = -\frac{3}{4}$ with $\Gamma = 0$
  renders this expansion complex for $x < 0$. This fits in nicely with
  the structure of expansions observed in the 0B case,
  reviewed\footnote{Recall that the trivial solution with $r(x) = 0$
    in that case did not have a real deformation for $q \neq 0$. The
    $v_3$ class is exactly the analogue of this trivial solution.} in
  section~\ref{sec:0B}.
\item $c = \Gamma$ causes these expansions to vanish (we got a first hint of this
  point in section~\ref{sec:painleve} where the second copy of
  Painlev\'e~IV has $\beta=0$.)
\item This class of expansions exists for all $n$, but is real as an
  $x<0$ expansion only for $n=2$~mod~$4$.

\end{itemize}
\subsubsection{Class 4}
\begin{itemize}
\item The $v_4$ class of expansions has not, to our knowledge, made a
  previous appearance in the literature, as it does not appear until
  encountering the DWW system.
\item $v_4$ contains powers of $c$ consistent with those of a
  parameter counting branes. This remains true for any value of
  $\Gamma$.
\item $v_4$ contains powers of $\Gamma$ consistent with those of a
  parameter counting fluxes. However, for arbitrary $c$, $g_s$ appears
  with odd powers, inconsistent with our requirements for a
  description of fluxes, as mentioned above. This problem is avoided
  if we set $c=0$ since this forces the odd powers of $g_s$ to vanish.
\item The direction of $v_4$ is not immediately determined by the
  consistency conditions we have used so far. Compatibility with the
  type~0 theories at $n=2$ requires $g_2=1$ which renders $v_4$ real
  for $x \rightarrow -\infty$. On the other hand, compatibility with
  the type~0 theories at $n=4$ requires $g_4=-\frac{3}{4}$ which
  renders $v_4$ real for $x \rightarrow +\infty$.
\item We will later provide evidence in favor of $v_4$ existing for $x
  > 0$.
\item In general, for the type~0 choices~\reef{tn} for $g_n$, $v_4$
  remains real for $x>0$ when $n = 0$ mod $4$ and becomes complex when
  $n = 2$ mod $4$.
\item In the special case $n=2$, $v_4$ can be made real by setting $c=0$.

\end{itemize}

\subsubsection{Class 5}
\begin{itemize}
\item $v_5$ contains powers of $\Gamma$ consistent with those of  a parameter
  counting branes. This is true only for $c=0$.
\item $v_5$ contains powers of $c$ consistent with those of  a parameter
  counting branes. This is true only for $\Gamma=0$.
\item As for $v_3$, we notice that associating one boundary to each
  factor of $c$ \emph{and} $\Gamma$ also produces a consistent
  worldsheet expansion. We might speculate that, in general, these
  expansions might capture $c$ and $\Gamma$ simultaneously counting
  branes.
\item These expansions do not exist for $n<3$.
\item With $\Gamma = 0$ and $g_4 =
  -\frac{3}{4}$ this class reduces to the $x<0$ brane--expansion in
  $c$ seen for the 0B theory at $n=4$ in equation
  (\ref{0Bexpnsm=4}). (Recall that there $q=-c$.)
\item These are the non--trivial broken--symmetry solutions obtained in
  the 0B theory as one increases $n$. We reviewed this at the end of
  section~\ref{sec:0B}.
\item As $n$ increases, further expansions in this class arise for
  every odd $n$, which we generically label $v_{i \geq 5}$. These are
  distinguished by the different values of $a$ in~\reef{classes}, but
  since their behavior is identical for our purposes we often group
  them together.
\item For odd $n$, reality imposes no restrictions on the direction of
  $v_{i \geq 5}$.
\item For $n=4$, reality requires that we fix $v_5$ to be an $x
  \rightarrow -\infty$ expansion. For the subsequent even $n$, some of
  the $v_{i \geq 5}$ are real for $+x$, while the remaining are real
  for~$-x$.
\end{itemize}

\subsection{The Structure at  Higher $n$}
We can extrapolate the pattern observed for the first few $n$ and make
predictions for the structures that should appear at higher $n$. The
first observation is that there are $(n+1)^2$ expansions in all
(taking into account the various expansions related by symmetries in
each class) at each $n$. The counting can be broken down as follows.

\begin{table}[!h!!]
\begin{center}
\begin{tabular}[t]{|c|c|c|c|c|c|c|c|r|}
\hline
&$v_2$&$v_4$&$v_1$&$v_3$&$v_5$&$v_6$&$v_7$&${\rm total}$\\
\hline
$n=1$&1&1&&2&&&&4\\
\hline
$n=2$&1&2&2&4&&&&9\\
\hline
$n=3$&1&3&&6&6&&&16\\
\hline
$n=4$&1&4&4&8&8&&&25\\
\hline
$n=5$&1&5&&10&10&10&&36\\
\hline
$n=6$&1&6&6&12&12&12&&49\\
\hline
$n=7$&1&7&&14&14&14&14&64\\
\hline
$n=8$&1&8&8&16&16&16&16&81\\
\hline
\end{tabular}
\caption{\small The number and types of expansion classes for $v(x)$,
  as a solution to the string equations~\reef{DWWstring1}
  and~\reef{DWWstring2}, with increasing $n$ from 1 to 8. See text
  for further discussion.}
\label{table:expansions}
\end{center}
\end{table}
Class 2 has exactly one member for each $n$, while Class 1 and Class 4
each have $n$ members. (Recall that Class 1 only exists for even $n$.)
As previously mentioned, for every odd $n$, new expansions in Class 5
(the $v_{i\geq 5}$) appear. These reduce to the $x<0$ broken symmetry
expansions of the 0B theory (for $\Gamma=0$ and $g_n$ in
equation~\reef{tn}) when $n$ is even. These expansion classes each
contain $2n$ members.  The appearance of these new expansions is
consistent with the counting provided in ref.\cite{Klebanov:2003wg}
for the 0B expansions, as reviewed at the end of section~\ref{sec:0B}.
The counting is tabulated in Table~\ref{table:expansions}. All
together, we see that for odd $n$, adding across the rows gives a
total of $1 + n + \frac{n+1}{2} \cdot 2n = (n+1)^2$ expansions, while
for even $n$ we get $1 + n + n + \frac{n}{2} \cdot 2n = (n + 1)^2$
expansions.

\section{An Organizing Square}
\label{sec:square}
To construct a full solution for $v(x)$, we need to specify its
behaviour in the two asymptotic directions, positive and negative
$x$. Consider the example of type~0A discussed in
section~\ref{sec:0A}. The string equation at a given $k$ was shown to
have a solution connecting these two perturbative regimes with a full
non--perturbative completion, plotted for $k=2$ in
figure~\ref{fig:plot}. (Recall that $z\propto x$ and $w=-v$.) This
solution is in fact made of two expansions, $v_1(x)$ for the positive
$x$ regime and $v_2(x)$ for negative $x$. The parameter $c$ is frozen
to $-\frac12$ in this case, leaving the parameter $\Gamma$ to count
D--branes at $+x$ and fluxes at $-x$, as described
in~\ref{sec:general}.

This is the organizing scheme we follow in order to construct more
theories, with type~0B being another working example, this time with
$\Gamma=0$ and using $v_1$ for positive $x$ and
$v_3,v_5$, or the higher $v_i$ (not $v_4$) for negative $x$, as
already discussed.

In constructing new theories, matching perturbative expansions
does not guarantee that a full non--perturbative solution exists with
the desired properties. Further work is needed, using both analytic
and numerical techniques, in order to demonstrate the non--perturbative
existence of the proposed theories. This is the subject of our
companion paper, where we find several non--perturbative solutions
numerically, and present analytical arguments in favour of several new
non--perturbatively complete theories. For the rest of this paper, our
analysis will be concerned with the various perturbative regimes that
appear from our DWW string equations.

\begin{figure}[ht]
  \begin{center}
 \includegraphics[width=100mm]{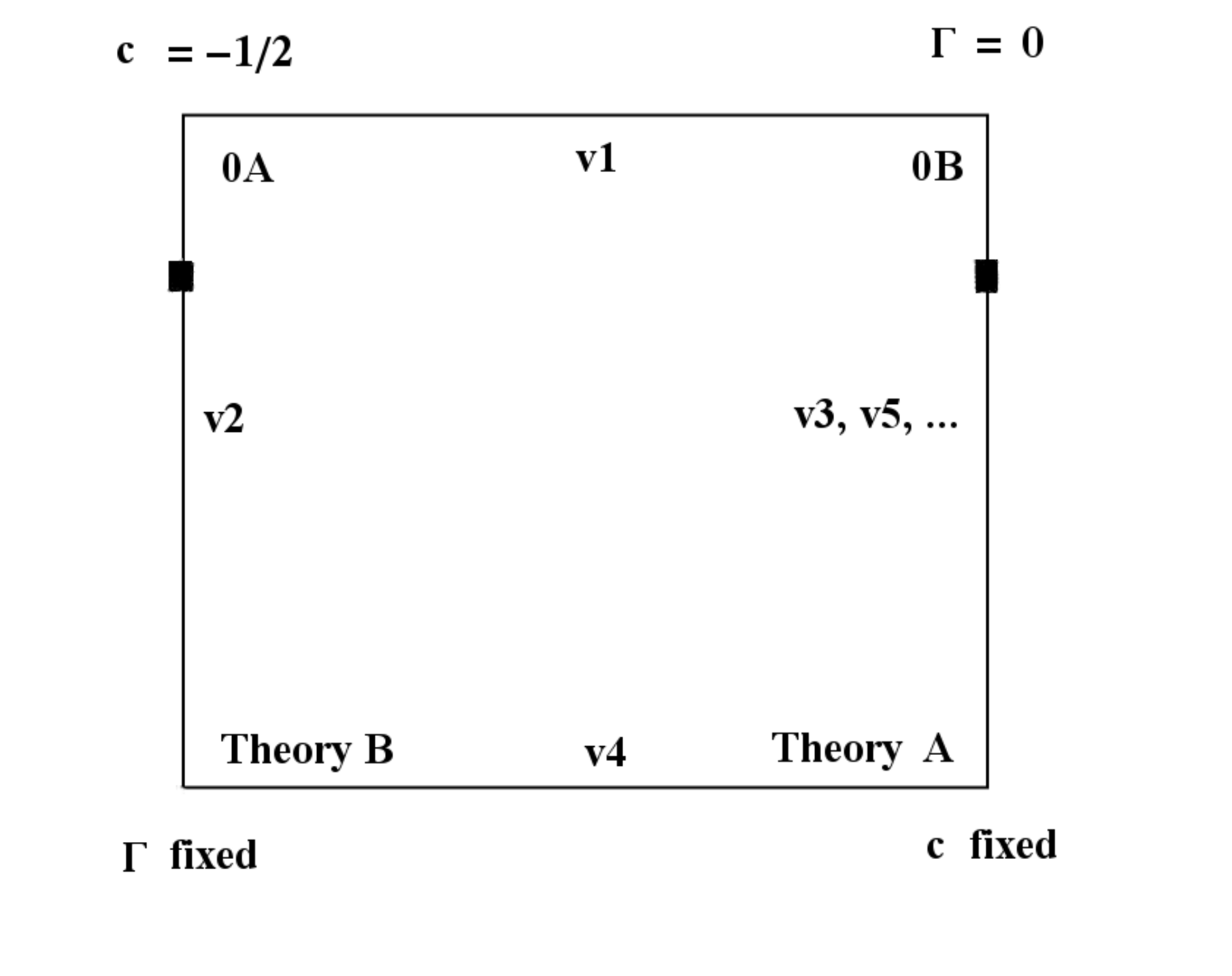}\\
 \caption{\small DWW Expansions forming a square. See text for
   explanation.}\label{square1}
 \end{center}
\end{figure}

Much of this structure can be organized neatly into the shape of a
square, with the string theory special points we know so far at two of
the corners.  The $v_1$ and $v_2$ pair form two edges with the type~0A
(with $c=-\frac12$) theory where they join. Then $v_3$ (at $n = 2$)
(or $v_5$ at $n = 4$, and so on) make another edge, with type~0B
($\Gamma=0$) at the corner where that edge meets the $v_1$ edge. See
figure~\ref{square1}.

Using our observations from section~\reef{sec:general}, we conjecture
that $v_4$ and $v_2$ form a physical pair when $\Gamma$ is fixed, with
$c$ counting either branes of fluxes in the perturbative
regimes. Similarly, $v_4$ with $v_3$ (or $v_{i \geq 5}$) may form
physical pair for fixed $c$, with $\Gamma$ counting fluxes or
branes. Since $v_2, v_3$, and $v_{i \geq 5}$ appeared as $x<0$
expansions, it is natural to fix the direction of $v_4$ to be $+x$. It
fits elegantly at the bottom of the square, at least when $n = 0 \mod
4$, ({\it i.e.,} when $v_4$ is real for positive $x$).

We summarize all of this in figure~\ref{square1}. The special points in
parameter space with $c^2 = \Gamma^2$ or $c=\Gamma$, where $v_2$ and $v_3$ vanish,
are represented by the dark squares on the vertical edges.

This way of organizing things immediately suggests that there are two
new special points, corresponding to the lower two corners of the
square, and we've called them Theory~A and Theory~B. We will need to
determine what the special values of $c$ and $\Gamma$ might be for
these corners, and the nature of the new theories. (The special values
$c=0$ and $\Gamma^2 = \frac{1}{4}$, complementary to the known values
for the type~0 theories, are suggestive, but so far this is a
guess. We will find several pieces of evidence to support this
suggestion in later sections.)

\begin{figure}[ht]
  \begin{center}
  \includegraphics[width=105mm]{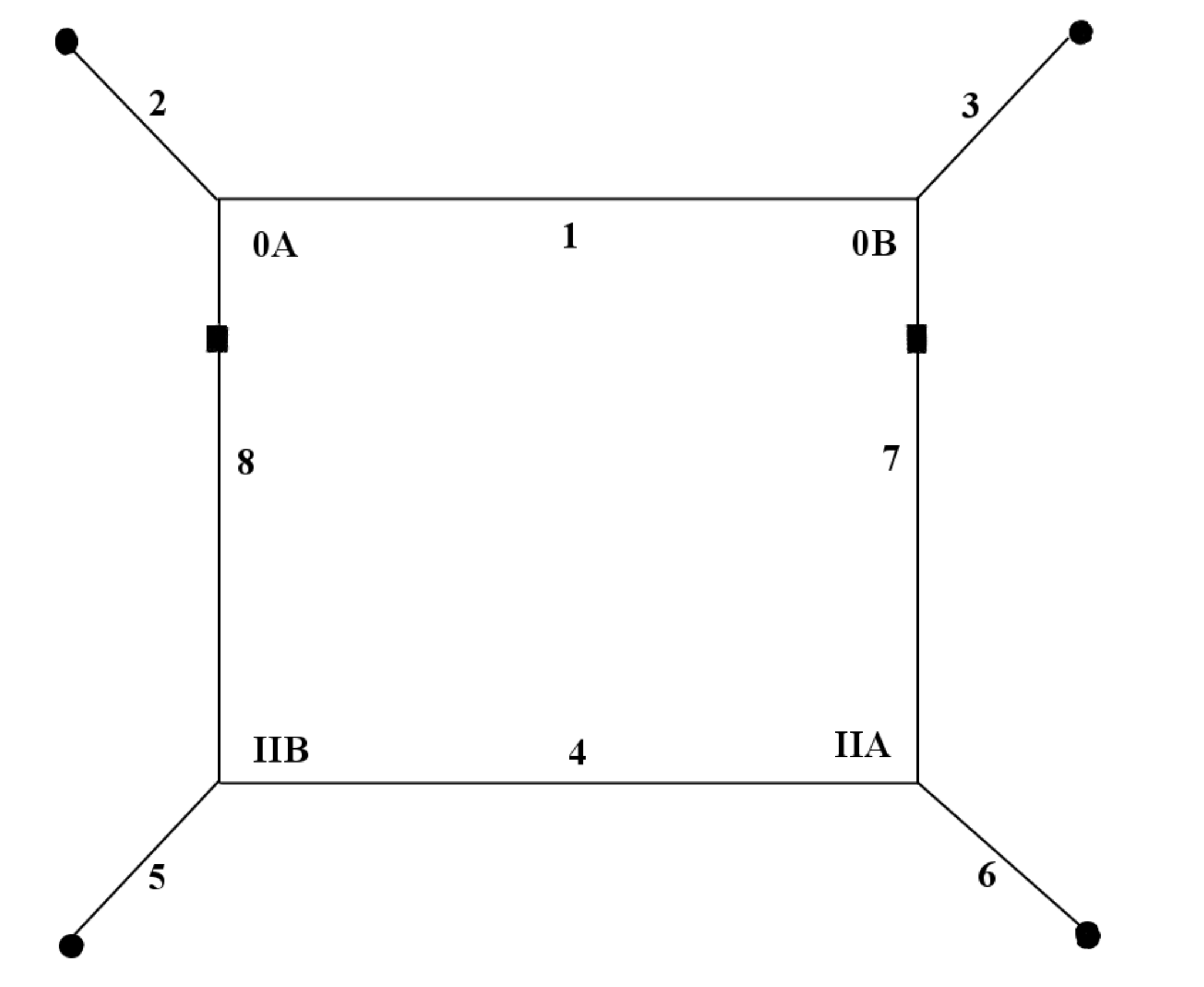}\\
  \caption{\small The moduli space of two--dimensional string
    theories\cite{Seiberg:2005bx}. The four corners of the square
    represent the four string theories 0B, 0A, IIB and IIA. The lines
    labelled~1--8 represent different compactifications. The points on
    each line represent compactifications with different radii
    $R$. Lines~1, 4, 7 and 8 interpolate between different
    non--compact theories as $R$ varies between $0$ and~$\infty$.  The
    points marked with black squares on lines 7 and 8 represent the
    non--critical superstrings of ref.\cite{Kutasov:1990ua}. See
    ref.\cite{Seiberg:2005bx} for discussion of other features of the
    diagram.}\label{fig3}
  \end{center}
\end{figure}

The lines connecting the special points are not (at this stage) to be
taken too literally, since we do not have a clear statement of the
nature of the theory (stringy or not) away from the special
points. However, the structure is highly suggestive, and reminiscent
of the square discovered in ref.\cite{Seiberg:2005bx} organizing the
moduli space of ${\hat c}=1$ strings. We reproduce it here in
figure~\ref{fig3}. ${\hat c}=1$ strings are fully two dimensional,
having in addition to the Liouville direction $\phi$ an extra
direction $X$. This direction can be compactified on a circle, and in
the square the lines represent values of radii varying between 0 and
$\infty$.  The relations between theories then arise as a result of
T--dualities (horizontally) possibly combined with discrete twists by
discrete fermionic symmetry operations (vertically).

In the case under study here, there is generically no compact circle,
and so the analogy is limited, but it is possible that it is not
entirely coincidental that a square emerges. Two dimensional string
theories can descend to ${\hat c}<1$
theories by Renormalization Group flow\cite{Gross:1990ub,Hsu:1992cm},
and so an organizing square at ${\hat c}=1$ may well leave an imprint
at ${\hat c}<1$ that still is an organizing square. The fact that
there are two special points on the vertical lines on the ${\hat c}=1$
square that match our $c^2=\Gamma^2$ and $c=\Gamma$ points is suggestive.

Inspired by the similarity between our square and that of
ref.\cite{Seiberg:2005bx}, we explored whether the two unknown
theories at the bottom of our square could actually be type~IIA and
type~IIB string theories, coupled to superconformal minimal models.
We present the details of our explorations in section~\ref{sec:new}.

\section{A Search for New Theories}
\label{sec:new}

So far, we have demonstrated that the DWW hierarchy has an extremely
rich structure of asymptotic expansions that naturally contain both
the 0A and 0B string theories coupled to the $(A,A)$ $(2,4k)$
superconformal minimal models. We've also pointed out that there are
new expansions that seem to have perfectly stringy interpretations, in
terms of backgrounds containing D--branes or fluxes once either $c$ or
$\Gamma$ has been fixed.  It is natural to wonder whether a sensible
interpretation as string theories coupled to some matter minimal
(super--minimal) models can be given to these new expansions. We will
find some success with this for some corners of parameter space.

The square in figure~\ref{fig3} motivates the conjecture that the new
string theories are type~II string theories coupled to some
superconformal minimal models.  This has some physical motivation
since type II string theories can be obtained as twisted orbifolds of
type 0 theories. Also, the $(A,D)$ series of superconformal minimal
models can be obtained as orbifolds of the $(A,A)$ series $via$ a
twist in the matter sector. In descending from ${\hat c}=1$ to ${\hat
  c}<1$, the remnants of the twisted T--dualities connecting the
type~0 and type~II sectors could well be a combination of these
orbifold actions.  In what follows, we argue that type~II string
theories coupled to $(A,D)$ $(4,4k-2)$ superconformal minimal models
are natural candidates for the physics encoded by the new special
points of our string equations.

One method of partially checking which theories are being captured by
our asymptotic expansions is to compare the (putative) torus
contributions (terms at order~$g_s^0$ in the free energy) with a
continuum calculation ({\it i.e.,} results of a traditional
world--sheet string one--loop computation) for these models. Such a
comparison will enable us to specialize to various points in parameter
space and provide further consistency checks to determine the exact
underlying models.

\subsection{The $g_s^0$ terms}
We begin by listing  the terms that appear at order $g_s^0$ in
the expansion for the free energy for each class of the expansions
studied in section~\ref{sec:beyond}, for all $n$:
\begin{eqnarray}\label{DWWtorusterms}
v_1(x)&:& \frac{n+1}{12n}-\frac{c^2}{n}+\frac{\Gamma^2}{n} \quad ,\nonumber\\
v_2(x)&:& c^2-\Gamma^2 \quad ,\nonumber\\
v_3(x)&:& \frac{n-1}{2n}c^2-\frac{n+1}{n}c\Gamma+\frac{n+3}{2n}\Gamma^2 \quad ,\\
v_4(x)&:& \frac{n+1}{6n}-\frac{c^2}{n}-\frac{2n+1}{n}\Gamma^2 \quad ,\nonumber\\
v_i(x)&:& \frac{n+1}{12n}-\frac{c^2}{n}+\frac{\Gamma^2}{n} \quad , \nonumber  \quad (i \geq 5)\ .
\end{eqnarray}
Notice that the torus term in the expansion classes labeled $v_i$ is
identical to that in $v_1$. This is a generalization of the curious
observation that was made in ref.\cite{Klebanov:2003wg}, that the
symmetry breaking solutions (for $x<0$) of the 0B theory have the same
torus terms (for $x>0$) as those of the 0B theory.

In each case, these terms are at order $x^{-2}$ in the expansion, and
multiplied by $\nu^2$. Following {\it e.g.,}
equations~\reef{0AFreeEnergy} and~\reef{0BFreeEnergy}, the free energy
is obtained by integrating twice and dividing by $\nu^2$, yielding the
same terms above multiplied by $\ln(|x|)$, which is part of a standard
Liouville theory volume factor that is common to everything we will do
at this order at perturbation theory\footnote{The Liouville direction
  $\phi$ is effectively a box of volume $V_{\rm
    L}=-\ln(|\mu|/\Lambda)/\alpha_{\rm min}$ where $\alpha_{\rm min}$
  is the Liouville dressing of the lowest dimension operator in the
  theory (see {\it e.g.,}
  refs.\cite{Ginsparg:1993is,DiFrancesco:1993nw} for a review). In a
  unitary theory, $\mu$ is the cosmological constant, the coefficient
  of the puncture operator, which measures worldsheet area. Recall
  that the dilaton and hence the local string coupling increases with
  $\phi$, and so there is a natural cutoff at the point where
  perturbation theory begins to break down, denoted
  $\Lambda$. Standard conventions are to choose a scale such that
  $\Lambda$ is unity and we will write $\mu=x$ in much of what
  follows, differing slightly from our notation in {\it e.g.,}
  equation~\reef{0AFreeEnergy}.}.

\subsection{The Continuum Partition Functions}\label{continuum}
We now present several continuum partition functions in the even spin
structures sector for both type~0 and type~II theories coupled to
$(A,A)$ and $(A,D)$ modular invariants. Ref.~\cite{Saleur}, presents
the modular invariant partition functions in the even spin structures
$(-,-)$, $(-,+)$ and $(+,-)$ for all the ${\cal{N}} = 1$
superconformal minimal models, as classified in
ref.~\cite{Cappelli}. Ref.~\cite{Bershadsky:1991zs} combines these
results with Liouville theory to compute some of the string theory
partition functions and we follow their methods to present the type~II
expressions that we suggest at the end of this section.

\subsubsection{The Type 0 Theories}

{$\bullet$} The $(A_{p-1}, A_{q-1})$ modular invariants. \\
The contribution of the {\it even} spin structures to the genus one
path integral for the $(A_{p-1}, A_{q-1})$ superconformal minimal
models coupled to supergravity has been calculated in
ref.\cite{Bershadsky:1991zs}:
\begin{eqnarray}\label{PFnAAmins1}
Z^{(A,A)}_{\rm even} &=& - \frac{1}{16} \frac{(p - 1)(q - 1)}{(p + q - 1)} \ln |x| \quad , \quad (p,q \quad {\rm odd})\\
\label{PFnAAmins2}
Z^{(A,A)}_{\rm even} &=& - \frac{1}{16} \frac{(p - 1)(q - 1) + 1}{(p + q - 2)} \ln |x| \quad . \quad (p,q \quad {\rm even})
\end{eqnarray}
\\
{$\bullet$} The $(A_{p-1}, D_{q/2 + 1})$ modular invariants.\\
The superconformal minimal model partition functions may be written in terms of
the partition functions for fields on a circle at special radii, as
shown in ref.\cite{Saleur}.  These can be combined with partition
functions for affinized compact circle theories to yield the desired
one--loop string theory expressions\cite{Bershadsky:1991zs}. Using
this technique, it is easy to show that the partition functions in the
even spin structures for the $(A_{p-1}, D_{q/2 + 1})$ modular
invariants are:
\begin{eqnarray}\label{PFnsADmins1}
Z^{(A,D)}_{\rm even} &=& - \frac{1}{64} \frac{(3p - 4)(q + 2)}{(p + q - 2)} \ln |x|\ , \quad (q = 2 \mod 4)\ ,\\
\label{PFnsADmins2}
Z^{(A,D)}_{\rm even} &=& - \frac{1}{32} \frac{(p - 2)(q + 3) + 2}{(p + q - 2)} \ln |x|\ ,  \quad (q = 0 \mod 4)\ .
\end{eqnarray}

\subsubsection{The Type II Theories}\label{sec:type-ii-z-even}
By analogy with the previous section, similar procedures can be used
to propose partition functions in the even spin structures for
superconformal minimal models coupled to the type II string theories, using as
starting point the partition functions for the corresponding circle
theories given in ref.\cite{Seiberg:2005bx}. Our results are:
\\
\newline
{$\bullet$} The $(A_{p-1}, A_{q-1})$ modular invariants. \\
\begin{eqnarray}\label{PFnAAminstypeII1}
\tilde{Z}^{(A,A)}_{\rm even} &=&  \frac{1}{32} \frac{(p - 1)(q - 1)}{(p + q - 1)} \ln |x| \quad , \quad (p,q \quad {\rm odd})\\
\label{PFnAAminstypeII2}
\tilde{Z}^{(A,A)}_{\rm even} &=&  \frac{1}{32} \frac{(p - 1)(q - 1) + 1}{(p + q - 2)} \ln |x| \quad . \quad (p,q \quad {\rm even})
\end{eqnarray}
{$\bullet$}  The $(A_{p-1}, D_{q/2 + 1})$ modular invariants.\\
\begin{eqnarray}\label{PFnsADminstypeII1}
\tilde{Z}^{(A,D)}_{\rm even} &=&  \frac{1}{64} \frac{p(q + 2)}{(p + q - 2)} \ln |x|\ , \quad (q = 2 \mod 4)\ ,\\
\label{PFnsADminstypeII2}
\tilde{Z}^{(A,D)}_{\rm even} &=&  \frac{1}{64} \frac{p(q + 2) - 4}{(p + q - 2)} \ln |x|\ ,  \quad (q = 0 \mod 4)\ .
\end{eqnarray}
While this is a natural extension of the definitions for bosonic
strings and type~0 strings {\it via}   combinations circle partition
functions, as we have already stated, an independent direct definition
of the type~II strings coupled to minimal models (explicitly coupling
to super--Liouville and defining the appropriate GSO projection) would
be desirable. This method only produces $Z_{\rm even}$, whereas a
direct definition would give explicit expressions for type~IIA and
type~IIB.

\subsection{Searching for Special Values of Parameters}\label{secSpecial}

\subsubsection{Comparison of Torus Terms --- The Known}
Before proceeding to the general story, we briefly review the
comparison\cite{Klebanov:2003wg} between the torus terms supplied by
the asymptotic expansions and the continuum calculations for the
type~0 theories coupled to the $(2,4k)$ (A,A) series of superconformal
minimal models.

For the $(2,4k)$ series, equation~\reef{PFnAAmins2} becomes
\begin{equation}\label{PFn2,4kmins}
Z_{\rm even}^{(2,4k)} = \frac{1}{2}\left(Z_{\rm 0A}(x) + Z_{\rm 0B}(x)\right) = -\frac{1}{16} \ln |x| \quad .
\end{equation}
Let us compare this with the results we have from our string equations
presented in section~\ref{sec:type0}. The torus terms in each
direction are as listed below:
\begin{eqnarray}\label{PFns0A0B}
Z_{\rm 0A} &=& -\frac{k-1}{24k} \ln |x|\ , \qquad Z_{\rm 0B} = -\frac{2k+1}{24k} \ln |x|\ , \qquad (x > 0) \ ,\nonumber\\
Z_{\rm 0A} &=& -\frac{1}{8} \ln |x|\ , \hskip0.75cm \qquad Z_{\rm 0B} = 0  \ , \qquad \hskip2.35cm (x < 0) \ .
\end{eqnarray}
These can be read off from the expansions given in
sections~\ref{sec:0A} and~\ref{sec:0B}, or alternatively by starting
with our DWW string equations and expansions given in
section~\ref{sec:beyond}, and specializing to either $c=-\frac12$
(type 0A) or $\Gamma=0$ (type 0B). It follows from the torus
terms~\reef{PFns0A0B} that
\begin{equation}\label{PFneven2,4k}
Z_{\rm even}^{(2,4k)} = \frac{1}{2}\left(Z_{\rm 0A}(x) + Z_{\rm 0B}(x)\right) = -\frac{1}{16} \ln |x| \quad ,
\end{equation}
for either sign of $x$, in agreement with the worldsheet computation
above. As argued in ref.\cite{Klebanov:2003wg}, this is strong
evidence that indeed these asymptotic expansions represent the
$(2,4k)$ super--minimal models coupled to supergravity.

Let us try to systematize the above procedure in the context of the
DWW string equations. Recall that specializing to the~0A~(0B) theory
requires $c = -\frac{1}{2}$ ($\Gamma = 0$). We seek to discover which
properties of our one--loop partition functions are general conditions
that produce these values.

To this end, turn again to the square of figure~\ref{square1}. It
suggests three possible theories: Theory $B$, which has $\Gamma$ fixed
and $v_2$ governing the $x\rightarrow -\infty$ asymptotics and $v_4$
governing the $x\rightarrow+\infty$ asymptotics (for which we
introduce the notation $(v_2|v_4)$); $\widetilde{A}$, which has $c$
fixed and $(v_3|v_4)$ asymptotics; and $\widehat{A}$, which has $c$
fixed and $(v_{i\ge 5}|v_4)$ asymptotics.  We summarize these
possibilities and the resulting torus terms in table~\ref{table:Torus
  terms}. These terms are obtained from~\reef{DWWtorusterms} by
eliminating $c$ or $\Gamma$ whenever its appearance represents an
insertion of a worldsheet boundary or a flux vertex operator, which
would change the topology.  So, {\it e.g.}, $\Gamma$ cannot appear in
the type 0A torus terms while~$c$ cannot appear for type 0B.

\begin{table}[!h!!]
\begin{center}
\begin{tabular}{|c|c|c|}
  \hline
  $$ & $x>0$ & $x<0$ \\
   \hline
  $Z_{\rm 0A}\phantom{\bigg(}$ &$a\left(\frac{n+1}{12n}-\frac{c^2}{n}\right)\  \quad\quad \left(v_1\right)$  & $a{c^2}\  \;\quad\quad\quad\quad \left(v_2\right)$ \\
  \hline
  $Z_{\rm 0B}\phantom{\bigg(} $& $b\left(\frac{n+1}{12n} + \frac{\Gamma^{2}}{n}\right)\  \quad\quad \left(v_1\right)$ & $b\left(\frac{n+3}{2n}\right)\Gamma^{2}\  \quad  \left(v_3\right)$ \\
  \hline
  $Z_{\rm \widetilde{A}}\phantom{\bigg(}$ & $\tilde{a}\left(\frac{n+1}{6n} - \frac{c^{2}}{n}\right)\  \quad\quad \left(v_4\right) $ & $\tilde{a}\left(\frac{n-1}{2n}\right)c^2\  \quad \left(v_3\right)$ \\
  \hline
  $Z_{\rm \widetilde{B}}\phantom{\bigg(}$& $\tilde{b}\left(\frac{n+1}{6n}-\frac{2n+1}{n}\Gamma^2\right)\  \quad \left(v_4\right)$ & $-\tilde{b}\Gamma^2\ \;\quad\quad  \quad \left(v_2\right)$ \\
  \hline
 $Z_{\rm \widehat{A}}\phantom{\bigg(}$ & $\hat{a}\left(\frac{n+1}{6n} - \frac{c^{2}}{n}\right)\ \quad \quad \left(v_4\right) $ &$\hat{a}\left(\frac{n+1}{12n}-\frac{c^2}{n}\right)\  \quad \left(v_{i\ge 5}\right)$ \\
   \hline
  $Z_{\rm \widehat{B}}\phantom{\bigg(}$& $\hat{b}\left(\frac{n+1}{6n}-\frac{2n+1}{n}\Gamma^2\right)\  \quad \left(v_4\right)$ & $-\hat{b}\Gamma^2\ \;\quad\quad  \quad \left(v_2\right)$ \\
  \hline
  
\end{tabular}
\caption{\small Unnormalized torus terms for string theories, before
  fixing parameters $c$ and $\Gamma$. We have also indicated from
  which expansion each term arises.}\label{table:Torus terms}
\end{center}
\end{table}

 As table~1 indicates, we have multiplied the torus terms by
unknown normalizations
$\{a,b\}$, $\{\tilde{a},\tilde{b}\}$ and $\{\hat{a},\hat{b}\}$, which we will later attempt to fix. Notice that we have distinguished two possibilities for $Z_B$: $Z_{\widetilde{B}}$ and $Z_{\widehat{B}}$, to match the ${\widetilde A}, {\widehat A}$ choices. We now construct the three possible $Z_{\mathrm{even}} = \frac{1}{2}\left(Z_A+Z_B\right)$ from these torus terms.\\
\newline 1. The known theories: 0A and 0B
\begin{eqnarray}\label{Zevenknown}
Z^{(\rm 0A, 0B)}_{\rm even} &=& \frac{1}{2}\left[\left(\frac{a+b}{12}\right)\left(\frac{n+1}{n}\right) - \frac{\left(ac^2 - b\Gamma^2\right)}{n}\right] \quad  \  \quad (x > 0) \nonumber\\
Z^{(\rm 0A, 0B)}_{\rm even} &=& \frac{1}{2}\left[a c^2 + \left(\frac{n+3}{2n}\right)b\Gamma^{2}\right]\;\;\quad\quad\quad\quad\quad \quad\quad\quad (x < 0)
\end{eqnarray}
2. The unknown theories: $\widetilde{\rm A}$ and $\widetilde{\rm B}$
\begin{eqnarray}\label{Zevenunknown2}
Z^{(\rm \widetilde{A}, \widetilde{B})}_{\rm even} &=& \frac{1}{2}\left[\left(\frac{\tilde{a} + \tilde{b}}{6}\right)\left(\frac{n+1}{n}\right) - \frac{1}{n}\tilde{a}c^2 - \left(\frac{2n+1}{n}\right)\tilde{b}\Gamma^2\right] \quad \quad (x > 0) \nonumber\\
Z^{(\rm \widetilde{A}, \widetilde{B})}_{\rm even} &=& \frac{1}{2}\left[\left(\frac{n-1}{2n}\right)\tilde{a}c^2 - \tilde{b}\Gamma^2\right] \quad\qquad\qquad\qquad\quad\qquad \qquad\quad (x < 0)
\end{eqnarray}
3. The unknown theories: $\widehat{\rm A}$ and $\widehat{\rm B}$
\begin{eqnarray}\label{Zevenunknown3}
Z^{(\rm \widehat{A}, \widehat{B})}_{\rm even} &=& \frac{1}{2}\left[\left(\frac{\hat{a} + \hat{b}}{6}\right)\left(\frac{n+1}{n}\right) - \frac{1}{n}\hat{a}c^2 - \left(\frac{2n+1}{n}\right)\hat{b}\Gamma^2\right] \quad ,  \quad (x > 0) \nonumber\\
Z^{(\rm \widehat{A}, \widehat{B})}_{\rm even} &=& \frac{1}{2}\left[\hat{a}\left(\frac{n+1}{12n}\right) - \frac{1}{n}\hat{a}c^2 - \hat{b}\Gamma^2\right] \qquad \qquad\qquad \qquad\quad \quad\quad (x < 0)
\end{eqnarray}

\subsubsection{The known theories: 0A and 0B}

We will now see what conditions are required to deduce the special
parameter values, $c = -\frac{1}{2}$ and $\Gamma = 0$, for type~0A
and~0B respectively.  We begin by observing that the continuum
partition functions listed in section~\ref{continuum} are independent
of the sign of $x$. Therefore, we will impose this as a constraint on
every $Z_{\rm even}$ that we construct,
\begin{equation}\label{Prop1}
\textrm{Condition 1:}\quad\quad 
Z_{\rm even}(x > 0) = Z_{\rm even}(x < 0)\quad. 
\qquad\qquad\quad
\end{equation}
We impose this condition by equating the two expressions in
equation~\reef{Zevenknown}. Interestingly, the $n$--dependence
factorizes completely leading to
\begin{equation}\label{RelncGammaAA}
ac^{2} + \frac{b}{2}\Gamma^{2} = \frac{a + b}{12} \quad .
\end{equation}
\indent While it is indeed possible for $c$ and $\Gamma$ to inherit
$n$ dependence from dependence on $t_n$, the simplest possibility is
that these parameters are independent of $n$. That the $n$ dependence
factors out of equation~\reef{RelncGammaAA} allows this simplicity to
be realized. This suggests that we also impose
\begin{equation}
\textrm{Condition 2:}\quad\quad 
c \textrm{ and } \Gamma \textrm{ are independent of } n . \\
\qquad\qquad\quad
\end{equation}
Substituting~\reef{RelncGammaAA} back into~\reef{Zevenknown} we
obtain,
\begin{equation}\label{Z0A0B}
Z^{(\rm 0A,0B)}_{\rm even} = \frac{a+b}{24}+\frac{3b}{4n}\Gamma^2
\end{equation}
Finally, we impose one more condition, 
\begin{equation}
\textrm{Condition 3:}\quad\quad 
Z_{\mathrm{even}} \textrm{ is independent of } n,
\qquad\qquad\quad
\end{equation}
which, together with Condition 2, forces us to conclude
 \begin{equation}\label{cGammaAAvalues}
c^{2} = \frac{a+b}{12a}\ , \quad \Gamma = 0 \quad \Rightarrow \quad Z^{(\rm 0A,0B)}_{\rm even} = \frac{a+b}{24}\; .
\end{equation}
\indent Condition 3 is motivated by the remarkable fact that
$Z_{\rm{even}}$ \emph{is} independent of $n$, which is nontrivial
given the form of equation~\reef{PFnAAmins2}, and also by the
observation that this constraint correctly produces the required
parameter values in this case. This is as far as we can go without
some extra information. Fortunately, for the type~0 theories, we
actually know $a= -\frac{1}{2}$ and $b = -1$: the factor of half is
because of the doubling of the free energy the 0B theory relative to
the 0A theory and the negative sign is because it is $-v$ which
defines the two--point function for these theories. (Recall the
observation at the end of section~\reef{sec:0Breduc}). Thus we see
that we obtain the correct parameter values $c=-\frac{1}{2}$ and
$\Gamma=0$ for the theories under consideration. Moreover, we
correctly obtain the known value\footnote{A subtle but important point
  should be mentioned here. We have used the torus term from $v_3$ to
  be the $x < 0$ contribution to the 0B theory leading us to $\Gamma =
  0$. This means that the $x < 0$ contribution to the 0B theory is
  actually zero. While $v_3$ is real as a $x < 0$ expansion for $n = 2
  \mod 4$, it is complex for $n = 0 \mod 4$. The resolution is that
  the zero contribution is to be understood as coming from the {\it
    trivial} $v = 0$ solution of the 0B theory obtained by setting $c
  = 0$ in the 0B string equations. It is interesting that our
  procedure using $v_3$ should give parameter values that are
  consistent with the {\it trivial} solution.} of
$Z_{\rm{even}}=-\frac{1}{16}$.

Next, we turn to exploring where our new conditions take us in
investigating the unknown corners.

\subsubsection{The unknown corners: $\widetilde{\rm A}$ and $\widetilde{\rm
    B}$}\label{sec:unknown corners}

Using equation~\reef{Zevenunknown2}, we see that Condition~1 gives,
\begin{equation}\label{RelncGammaAD}
\frac{\tilde{a}}{2}c^{2} + \tilde{b}\Gamma^{2} = \frac{\tilde{a} + \tilde{b}}{6} \quad \Rightarrow \quad Z^{(\rm \widetilde{A}, \widetilde{B})}_{\rm even}= -\frac{\tilde{a}+\tilde{b}}{12} +\tilde{a}c^2\left(\frac{1}{2}-\frac{1}{4n}\right)\, .
\end{equation}
Conditions~2 and~3 give,
\begin{equation}\label{GammaAD}
\Gamma^{2} = \frac{\tilde{a} + \tilde{b}}{6\tilde{b}} \quad \mathrm{and} \quad
  c = 0 \ ,
\end{equation} and so\footnote{This $c = 0$ result means that the
  contribution to the torus terms coming from $v_3$ is zero. In a
  sense, this result is the analog of what
  was seen for the type~0B theories. Our argument again is that the torus term
  for $x<0$ at the $(v_4|v_3)$ corner comes from the trivial $v=0$
  solution, analogous to the 0B case.} we have
\begin{equation}\label{ABtilde}
Z^{(\rm \widetilde{A}, \widetilde{B})}_{\rm even} = -\frac{\tilde{a} + \tilde{b}}{12} \quad .
\end{equation}
Remarkably enough, everything seems to work as before, except now we
obtain new values for our parameters and a new expression for $Z_{\rm
  even}$.  We have yet to make an numerical prediction for $Z_{\rm
  even}$ as we still must fix the normalizations. We will return to
this point in next section~\ref{sec:new-theories}.

\subsubsection{The unknown corners: $\widehat{\rm A}$ and $\widehat{\rm
    B}$}\label{sec:unknown-tilde}

Condition~1 gives:
\begin{equation}
\Gamma^2=\frac{\hat{a}+2\hat{b}}{12\hat{b}}\quad \Rightarrow \quad Z^{(\rm \widehat{A}, \widehat{B})}_{\rm even} = -\frac{\hat{b}}{12}+\frac{\hat{a}(1-12c^2)}{24n}
\end{equation}
Conditions~2 and~3 imply,
\begin{equation}
c^2=\frac{1}{12}, \quad\Gamma^2=\frac{\hat{a}+2\hat{b}}{12\hat{b}}\quad\Rightarrow\quad Z^{(\rm \widehat{A}, \widehat{B})}_{\rm even} = -\frac{\hat{b}}{12}
\end{equation}
This result is not as encouraging. $Z_{\rm even}$ does not depend on
$\hat{a}$ which implies that it does not depend on theory $\hat{A}$ at
all. We find this unacceptable and therefore conclude that
Conditions~2 and~3 are incompatible constraints in this case. We
nevertheless take seriously the possibility that theories like this
may exist, but with $n$--dependent parameters and an $n$--dependent
$Z_{\rm even}$. We will briefly consider this possibility in
section~\reef{sec:ndep}.

\subsection{The New Theories}\label{sec:new-theories}
In the above we have explored the crucial and quite constraining
assumption that $Z_{\rm even}$ is independent of the index $n =
2k$. Theories $\tilde{\rm A}$ and $\tilde{\rm B}$ seemed particularly
suited to produce such a $Z_{\rm even}$. In an effort to better
understand what these theories may be, we next ask which continuum
theories are capable of producing an $n$--independent $Z_{\rm
  even}$. Looking at the various partition functions listed in
section~\ref{continuum}, it is easy to see that only the following
choices give
rise to an $n$--independent partition function:\\
\newline 1. $Z^{(A,A)}_{\rm even}$ in equation~\reef{PFnAAmins2} with
$p=2$ and $q=4k$.\footnote{The two positive integers $p$ and $q$
  labeling the super--minimal models must obey: $q > p$; $q - p = 0
  \mod 2$; if both are odd, they are coprime and if both are even,
  then $p/2$ and $q/2$ must be coprime. There is also a standard
  restriction that if $p$ and $q$ are even, then $(q-p)/2$ must be
  odd. It follows that if $p=2$ then $q=4k$.} These theories are
type~0 string theories coupled to the $(2,4k)$ $(A,A)$ models and are
already described
at the known corners.\\
\newline 2. $Z^{(A,D)}_{\rm even}$ in equation~\reef{PFnsADmins1} with
$p=4$ and $q = 4k - 2$.\footnote{ Strictly speaking, $q = 4k \pm
  2$. The choice $q = 4k + 2$ would suggest that the $(4,6)$ $(A,D)$
  model exists at at $k=1$; however, all combinations of real
  expansions at $k=1$ have been exhausted in describing the type~0
  $(A,A)$ $(2,4)$ model. We therefore expect
  the $(4,6)$ model to appear at $k=2$ where more combinations of expansions exist since $v_4$ is real. Thus we choose $q=4k-2$ with $k\ge 2$.}  These theories are the type~0 strings theories coupled to the $(4, 4k-2)$ $(A,D)$ models with $Z^{(A,D)}_{\rm even} = - \frac{1}{8} \ln |x|$.\\
\newline 3. $\tilde{Z}^{(A,A)}_{\rm even}$ in
equation~\reef{PFnAAminstypeII2} with $p=2$ and $q=4k$. These are the
type II strings coupled to the $(2,4k)$ $(A,A)$
models with $\tilde{Z}^{(2,4k)}_{\rm even} = \frac{1}{32} \ln |x|$.\\
\newline 4. $\tilde{Z}^{(A,D)}_{\rm even}$ in
equation~\reef{PFnsADminstypeII2} with $p=2$ and $q=4k$. These are the
type II strings coupled to the $(2,4k)$ $(A,D)$ models with
$\tilde{Z}^{(2,4k)}_{\rm even} = \frac{1}{32} \ln |x|$. This is the
same partition function as case~$(3)$ above.
\\
\newline 5. $\tilde{Z}^{(A,D)}_{\rm even}$ in
equation~\reef{PFnsADminstypeII1} with $p=4$ and $q = 4k-2$.  These
theories are the type~II string theories coupled to the $(4,4k-2)$
$(A,D)$ models with
$\tilde{Z}^{(4,4k-2)}_{\rm even} = \frac{1}{16} \ln |x|$.\\
\newline We seek to identify $Z^{(\rm \widetilde{A},
  \widetilde{B})}_{\rm even} = -\frac{\tilde{a} + \tilde{b}}{12}$ with
one of the above partition functions, but to do so we will need some
way of determining $\tilde{a}$ and $\tilde{b}$. We consider the
possibility that the theories $\rm \widetilde{A}$ and $\rm
\widetilde{B}$ are type~IIA and type~IIB string theories,
respectively. As we will see, this will uniquely fix the
normalizations and allow us to identify the resulting $Z^{(\rm
  \widetilde{A}, \widetilde{B})}_{\rm even}$ as that of case~(5)
above, {\it i.e.} type~II string theories coupled to the $(4,4k-2)$
$(A,D)$ models.

\subsubsection{Type~II Theories Coupled to the Superconformal Minimal Models}
Pursuing the apparent similarity with the moduli space of ${\hat c}=1$
theories\cite{Seiberg:2005bx} a little further, (recall our discussion
in section~\ref{sec:square}, and figure~\ref{fig3}) we can study the
partition functions of those theories (compactified on a circle) to
get clues as to the possible relative normalizations. We list them
here\cite{Seiberg:2005bx}:
\begin{eqnarray}
{\rm 0A:} \qquad && \frac{Z}{V_L}=\frac{1}{12\sqrt{2}}\left(\frac{\sqrt{\alpha^\prime}}{R}+2\frac{R}{\sqrt{\alpha^\prime}}\right) \ , \nonumber \\
{\rm 0B:} \qquad && \frac{Z}{V_L}=\frac{1}{12\sqrt{2}}\left(2\frac{\sqrt{\alpha^\prime}}{R}+\frac{R}{\sqrt{\alpha^\prime}}\right)\ , \nonumber\\
{\rm IIA:} \qquad && \frac{Z}{V_L}=-\frac{1}{24\sqrt{2}}\left(2\frac{\sqrt{\alpha^\prime}}{R}+\frac{R}{\sqrt{\alpha^\prime}}\right)\ , \nonumber \\
{\rm IIB:} \qquad && \frac{Z}{V_L}=-\frac{1}{24\sqrt{2}}\left(\frac{\sqrt{\alpha^\prime}}{R}+2\frac{R}{\sqrt{\alpha^\prime}}\right)\ .
\end{eqnarray}
An important clue is to be found in the $1/R$ behaviour of each
theory. This is the physics of the field theory sector (Kaluza--Klein
states) that propagate on the circle, and the relative normalizations
are of interest to us. Type~0B has twice as much energy in this sector
as type~0A, and type~IIA has double that of type~IIB.  Now, as
expected, the type~0 theories exchange under the T--duality operation
$R\to\alpha^\prime/R$, as do the type~II theories. Now T--duality
vertically should take the type~II theories to the type~0 theories,
but this needs to be done on a circle twisted by $(-1)^{f_l}$. The
partition functions for IIA/B on such a circle is:
\begin{eqnarray}
{\rm IIA:} \qquad && \frac{Z}{V_L}=-\frac{1}{24\sqrt{2}}\left(\frac{\sqrt{\alpha^\prime}}{R}-\frac{R}{\sqrt{\alpha^\prime}}\right) \ , \nonumber \\
{\rm IIB:} \qquad && \frac{Z}{V_L}=-\frac{1}{12\sqrt{2}}\left(\frac{\sqrt{\alpha^\prime}}{R}-\frac{R}{\sqrt{\alpha^\prime}}\right)\ .
\end{eqnarray}
Notice that the field theory term of type~IIA is now actually half
that of type~IIB. Now we can compare the overall type~II normalization
to the type~0 by doing the vertical T--duality. Under
$R\to\alpha^\prime/R$ we see that type~IIB field theory term matches
that of the type~0A theory, while for type~IIA, the field theory term
is 1/4 that of the type~0B value.

In summary, we see that comparing squares would suggest that $Z_{\rm
  A}=Z_{\rm IIA}$ and $Z_{\rm B}=Z_{\rm IIB}$, and the chain of
relationships above gives the relative normalization of 0B and 0A as 2
to 1, while that of (twisted) IIB and IIA is 2 to 1, so ${\tilde
  b}=2{\tilde a}$. Finally, T--duality (with a twist) between 0A
and IIB gives them the same relative normalization, implying,
\begin{equation}
\tilde{b}=-\frac{1}{2}, \quad \tilde{a}=-\frac{1}{4}\; .
\end{equation}
In this case we get therefore that for our putative type~II theories,
\begin{equation}
Z^{(\rm \tilde{A}, \tilde{B})}_{\rm even}=\frac{1}{16}\ln |x| \; ,
\end{equation}
which is equal to that of the type~0 case, but with an opposite
sign. This is the same as the partition
function~\reef{PFnsADminstypeII1} for the type~II theories coupled to
the $(4,4k-2)$ $(A,D)$ super--minimal models.  Interestingly,
equation~\reef{GammaAD} implies that $\Gamma^2=1/4$ for IIB and $c=0$
for IIA, which nicely mirrors the values $c^2=1/4$ for 0A and
$\Gamma=0$ for 0B.

Further work is required to strengthen our conjecture that we have
here the type~IIA and~IIB theories coupled to the superconformal
minimal models. As stated, an explicit definition of ${\hat c}<1$
type~II strings (separately for types~A and~B, and not just our
suggestion for $Z_{\rm even}$ given in
section~\ref{sec:type-ii-z-even}) does not yet seem to exist in the
literature, and our attempts to directly define them so far are
incomplete. Note that we have assumed that the relative normalizations
of the partition functions that follow from ${\hat c}=1$ really
descend to the ${\hat c}<1$ case, and while reasonable, this needs to
be proven. The explicit $k$ dependence of the individual partition
functions that we get by taking the DWW one--loop expressions
seriously have a (so far) unilluminating form at positive $x$ that we
have not been able to check against a continuum computation. On the
other hand the negative $x$ result neatly mirrors the type~0 case. We
list the results here, and leave this 
line of investigation for the future\footnote{We also considered the
  possibility that our $\tilde{A}$ and $\tilde{B}$ theories are again
  type~0A and type~0B string theories, respectively, perhaps coupled
  to an $(A,D)$ modular invariant. This would again fix the relative
  normalization of the partition functions such that $a=-\frac{1}{2}$
  and $b=-1$, yielding $Z^{\rm (\widetilde{A},\widetilde{B})}_{\rm
    even} = \frac{1}{8} \ln |x| $, which unfortunately differs from
  the continuum partition function for the $(4, 4k-2)$ $(A,D)$
  super--minimal models by a sign. To match, we would have to assume
  that $+v$ encodes the partition function instead of $-v$, which
  contradicts our earlier establishment of the $(A,A)$ type~0
  points.}:
\begin{eqnarray}\label{PFnsIIAIIB}
Z_{\rm IIA}&=&-\frac{2k+1}{48 k}\ln|x|\ ,  \qquad Z_{\rm IIB}=\frac{8k+1}{48k}\ln|x|\ ,  \qquad (x>0)\ ,\nonumber \\
Z_{\rm IIA}&=&0\ , \qquad \hskip2.3cm   Z_{\rm IIB}=\frac18\ln|x|\ , \hskip0.9cm\qquad (x<0)\ .
\end{eqnarray}
This should be compared with the type~0 cases displayed in
equation~\reef{PFns0A0B}.

\subsubsection{$n$--dependence}\label{sec:ndep}
Our main successes so far have involved theories in which $Z_{\rm
  even}$ and the values for $\Gamma$ and $c$ have been independent of
$n$. While these are the simplest possibilities, there is little
reason to suspect that these theories are the only types amenable to
description by DWW. We observed in section~\ref{sec:unknown-tilde}
that there might be new theories, {\it e.g.,} the ones we called
$\widehat{A}$ and~$\widehat{B}$, whose description will most likely
require such $n$--dependence for $(\Gamma,c)$.  Unfortunately, the
availability of continuum calculations to compare to is limited, so
obtaining convincing evidence for any particular proposal is
difficult. Nevertheless, there is much more evidence to accumulate
from the differential equations themselves. We hope to report upon
this issue in later work.

We briefly mention one way that $n$--dependence might emerge
naturally. The $Z_{\rm even}$ that we have constructed from the torus
terms of our perturbation expansions depend on three variables: $n$,
$c$, and $\Gamma$. We must however impose sign independence which
reduces the total number of independent variables describing $Z_{\rm
  even}$ to two. In special cases the algebra conspires and we find
that $Z_{\rm even}$ actually depends on no variables at all. This
behavior is to be compared to $Z_{\rm even}$ as computed from the
continuum partition functions. For generic $p$ and $q$, $Z_{\rm even}$
is a function of two variables, but in special cases this dependence
completely vanishes. This motivates us to attempt to express for $n$,
$c$, and $\Gamma$ as functions of $p$ and $q$.

Equation~\reef{Z0A0B} indicates that $Z_{\rm even}^{\rm 0A, 0B}$ is
inversely proportional to $n$. The form of equation~\reef{PFnAAmins2} then
suggests that we take $n\sim p+q-2$. Since $q=4k=2n$ in this case, we
can predict
\begin{equation}
n=\frac{1}{2}(p+q-2)\ .
\end{equation}
Now equating $Z_{\rm even}^{\rm 0A, 0B}$ of equation~\reef{Z0A0B} with
the quantity in equation~\reef{PFnAAmins2} and using the relation
$c^2+\Gamma^2=\frac{1}{4}$, which follows from sign independence,
gives
\begin{equation}
\Gamma^2 =\frac{1}{12}(p-2)(q-2)\ .
\end{equation}
Note that the models already studied had $p=2$ which gave us
$\Gamma=0$. We can perform the same exercise for $Z^{(\rm \tilde{A},
  \tilde{B})}_{\rm even}$ which we have argued describes type II
strings coupled to $(A,D)$ $(4,4k-2)$ models. Again we have
$n=\frac{1}{2}(p+q-2)$, which holds for $p=4$ and
$q=4k-2=2n-2$. Equating $Z^{(\rm \tilde{A}, \tilde{B})}_{\rm even}$ of
equation~\reef{RelncGammaAD} with the quantity in
equation~\reef{PFnsADminstypeII1} and using $\frac{1}{4}c^2+\Gamma^2 =
\frac{1}{4}$ gives,
\begin{equation}
  c^2=-\frac{(p-4)(q-2)}{8(p+q-3)} \ .
\end{equation}
 For the models we've previously
considered, $p=4$, so we correctly reduce to the condition $c=0$. Note
here that we would run into problems if we considered $p>4$ because
then $c$ would become imaginary. 

\section{Discussion}
\label{sec:conclusion}
As discussed at length in the introduction, we find very significant
this rich framework into which we can embed so many string theories and
discover how they intertwine with each other.  We suspect that there
may be many more theories to be found in this framework, and that we
have only just begun to learn how to extract and identify the various
theories using the limited comparisons we can do to existing continuum
computations.

\begin{figure}[ht]
  \begin{center}
 \includegraphics[width=100mm]{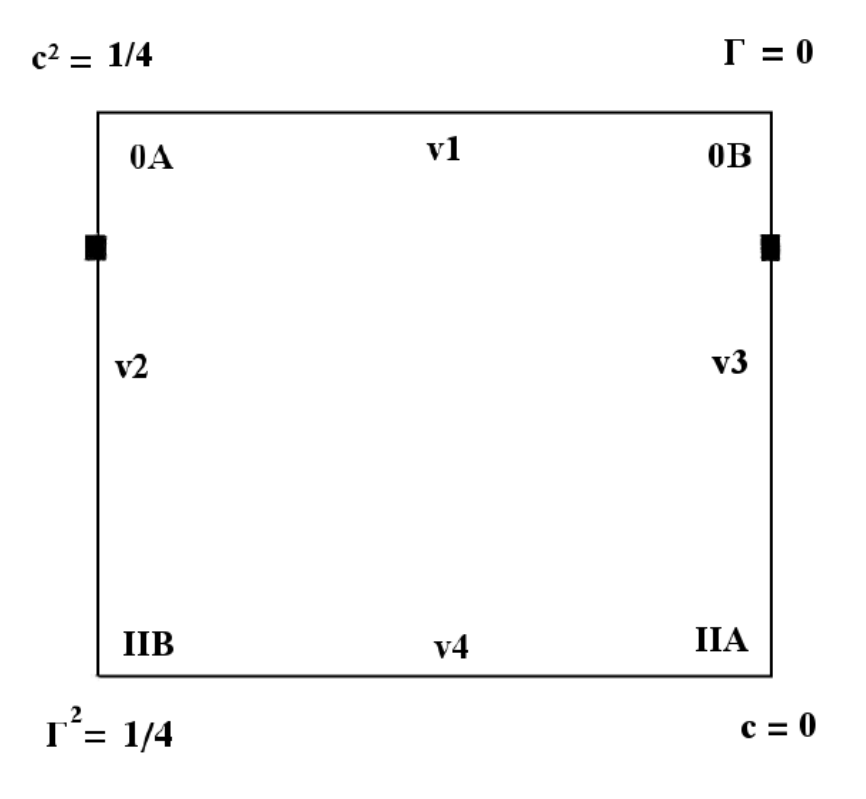}\\
 \caption{A family of string theories, forming a square. See text for
   details.}\label{square2}
 \end{center}
\end{figure}

We have been able to gather evidence for a square of connected
theories, firmly establishing the top corners as type~0A and~0B
theories coupled to the $(A,A)$ modular invariant $(2,4k)$ minimal
models, and finding several strong pieces of evidence that the special
points at the bottom are string theories, the
(non--supersymmetric\cite{Seiberg:2005bx}) type~IIA and~IIB theories
coupled to the $(A,D)$ modular invariant $(4,4k-2)$ minimal
models. See figure~\ref{square2}.

Notice that the $v_3$ side of the square is generically understood as
defining the theory when the number background branes, {\it i.e.,}
$c$ or $\Gamma$, is set to zero. This is the case for type~0B (already established
in ref.~\cite{Klebanov:2003wg}) and is inherited by the type~IIA
theory as well. Away from $c=0$ $v_3$ is complex for $k$ even, and one
of the symmetry breaking $v_i (i\geq5)$ may supply the $x<0$ physics
instead. (For $k$ odd, $v_3$ is real for $c\neq0$ and so may well
furnish the $x<0$ physics in those cases.)

It is also important to note that for $k$ odd, $v_4$ is no longer real
(with the exception of $c=0$ at $k=1$). This suggests that this regime
of perturbation theory at $x>0$ (containing fluxes for type~IIA and
branes for type~IIB) is ill--defined at those values of $k$, even
while the opposite regime at $x<0$ exists. This may suggest
non--perturbative instabilities for those values of $k$. An
instructive prototype of this possibility is the original bosonic
family of string theories defined non--perturbatively in terms of a
Painlev\'e~I hierarchy of string
equations\cite{Gross:1989vs,Brezin:1990rb,Douglas:1989ve}. At large
$x$, the leading behaviour for the two point function was $w(x)=x^k$
for the $k$th model, and for positive $x$ the physics was the
$(2,2k-1)$ conformal minimal models coupled to Liouville theory. For
even $k$, the the $x<0$ regime gives complex values, signalling the
non--perturbative problems. Something analogous may be going on here
for odd $k$ (except $k=1$, $c=0$) for the type~II $(A,D)$ $(4,4k-2)$
models. We will report further non--perturbative aspects in a
follow--up paper where we discuss numerical and analytic studies of
solutions that connect different perturbative regimes\cite{companion}.

There are many more interesting issues to understand. While we
compared to our proposal for a continuum definition of the even sector
partition function of the type~II theories, a direct continuum
definition of the type~IIA and~IIB string theories would be valuable
to have, in order to provide another check on our conjecture for the new
points. Whether or not they are type~II string theories, it is clear
that these new theories are of interest, and may (as already stated)
be only the first in a very large family of new theories that our
string equations define. They are all nicely interconnected with the
known type~0 theories, and have a rich non--perturbative sector. We
will explore much of this non--perturbative physics in a follow--up
paper\cite{companion}.

We have seen several signs that the organizing square is inherited
from the organizing square of theories seen at ${\hat c}=1$ in
ref.\cite{Seiberg:2005bx}. We suspect this inheritance arises
physically by RG flow, and it would be of value to explore this
further (there are bosonic investigations in the
literature\cite{Gross:1990ub,Hsu:1992cm}). In particular, the special
points with $\Gamma{=}\pm c$, where our partition functions vanish
(also suggested by the underlying Painlev\'e~IV structures we saw in
section~\ref{sec:painleve}, which are also worth understanding
better), may well be related to the supersymmetric points identified
in ref.\cite{Seiberg:2005bx}.

In summary, we have found a rich laboratory of solvable string theory
models with several non--perturbative connections between them by
realizing them as special points of a larger physical system, the
theory of dispersive water waves. In some respects it is an analogue
of what we would like to see in studies of M--theory. It will be
interesting to learn whether the larger framework of dispersive water
waves can yield any new insights about the non--perturbative nature of
string theories and related theories.

\section*{Acknowledgements}
This work was supported by the US Department of Energy. RI thanks
Tameem Albash, Nikolay Bobev, Arnab Kundu, Hubert Saleur, and Nicholas
Warner for useful conversations. CVJ thanks the Aspen Center for
Physics for hospitality while some of this work was carried out.


\providecommand{\href}[2]{#2}\begingroup\raggedright\endgroup

\end{document}